\documentclass[letterpaper,11pt]{article}
\usepackage[colorinlistoftodos,prependcaption,textsize=tiny]{todonotes}

\usepackage{xurl}
\usepackage{hyperref}

\usepackage{xcolor}
\usepackage{graphicx}
\usepackage{amsmath}
\usepackage{amsthm}
\usepackage{amssymb}
\usepackage{array}
\usepackage{bm}
\usepackage{hyperref}
\usepackage{geometry}[margin=1in]
\usepackage{bbm}
\usepackage{tcolorbox}
\usepackage{diagbox}
\usepackage{natbib}
\usepackage{siunitx}
\usepackage{booktabs, nicematrix, subcaption}

\def\Gauss{{ \mathrm{N}}}
\renewcommand{\epsilon}{\varepsilon}

\newcommand{\footremember}[2]{%
    \footnote{#2}
    \newcounter{#1}
    \setcounter{#1}{\value{footnote}}%
}
\newcommand{\footrecall}[1]{%
    \footnotemark[\value{#1}]%
} 

\newcommand{\datafield}[1]{\texttt{#1}}

\usepackage{siunitx}

\definecolor{darkread}{rgb}{0.7, 0, 0}

\newcommand{\dr}{\color{black}}

\title{State Space Modeling of Mortgage Default Rates under Natural Hazard Shocks}
\author{
    Samuel J. Eschker \footremember{purdue}{Purdue University -- Department of Statistics} \\ \texttt{seschker@purdue.edu}
    \and
    Antik Chakraborty \footrecall{purdue} 
    \thanks{Corresponding author; Postal address: 150 N.\ University St, West Lafayette, IN 47907 }\\ \texttt{antik015@purdue.edu} \and
    Melanie Gall \footremember{asu}{Arizona State University -- School of Public Affairs} \\ \texttt{melanie.gall@asu.edu}
    \and
    Peter Jevtic \footrecall{asu}\footremember{asumath}{Arizona State University -- School of Mathematical and Statistical Sciences} \\ \texttt{peter.jevtic@asu.edu}
    \and
    Jianxi Su \footrecall{purdue} \\ \texttt{jianxi@purdue.edu}
}
\date{}
\begin{document}

\maketitle

\begin{abstract}
Mortgage default rates, on the one hand, serves as a measure of economic health to support decision-making by insurance companies, and on the other hand, is a key risk factor in the asset-liability management (ALM) practice, as mortgage-related assets constitute a significant portion of insurers’ investment portfolios.  This paper studies the relationship between economic losses due to natural hazards and mortgage default rates. The topic is greatly relevant to the insurance industry, as excessive insurance losses from natural hazards can lead to a surge in mortgage defaults, creating compounded challenges for insurers. To this end, we apply a state-space modeling (SSM) approach to decouple the effect of natural hazard losses on mortgage default rates, after controlling for other economic determinants through the inclusion of latent variables.  Moreover, we consider a sliced variant of the classical SSM to capture the subtle relationship that only emerges when natural hazard losses are sufficiently high. Our model verifies the significance of this relationship and provides insights into how natural hazard losses manifest as increased mortgage default rates.

\end{abstract}
\noindent%
{\it Keywords:}  Bayesian inference; Latent factors; Loss modeling; Sliced method; Dependence

\section{Introduction}
The mortgage default rate refers to the percentage of mortgage loans that are delinquent, typically defined as past due by a specified period. In this paper, we define a 90-day delinquency as default.  This is consistent with the classification by the US Consumer Financial Protection Bureau, which categorizes such cases as serious delinquencies reflecting severe economic distress among borrowers. In general, studying mortgage default rates can provide useful insights into the overall health of the economy, which in turn can help insurance companies understand and anticipate policyholder behaviors, including demand for new policies and lapse tendencies.

%[Why is modeling and predicting mortgage default rate important for the insurance industry?] 
Beyond serving as a measure of economic health,  modeling and understanding the determinants of mortgage default rates is important to the insurance industry for several more direct and practical reasons. %First, mortgage insurance represents a important sector of the insurance industry. Changes in borrower default behavior on loans are a key risk factor influencing the management of mortgage insurance policies. Second, 
Specifically, mortgage-backed securities (MBS) constitute a significant portion of insurance companies' investment portfolios. According to a recent report published by the NAIC Capital Markets Bureau, as of year-end 2023, mortgage-related investments accounted for approximately 9\% of the investment assets held by insurance companies \citep{NAIC2023}. These assets represent the third-largest portion of insurers' investments, following bonds (60.8\%) and common stocks (13.9\%). 
The inclusion of MBS in investment portfolios inherently links mortgage defaults to insurance companies' asset-liability management (ALM) practices. Adverse changes in mortgage default rates can disrupt the predicted cash flows from mortgage-related assets, thus triggering mismatches between an insurer's asset inflows and liability outflows. Increases in mortgage default rates can also cause valuation declines in MBS, increased duration risk, and challenges in maintaining liquidity and regulatory capital adequacy.  

In addition, changes in borrower default behavior on loans are a key risk factor influencing the management of mortgage insurance business, which represent an important segment of the insurance industry.%In addition to the aforementioned more direct impacts, mortgage default rates can provide useful insights into the health of the economy, which in turn can help insurance companies understand and anticipate policyholder behaviors such as demand and lapse behaviors.

Modeling the mortgage default rates is becoming more important yet challenging in the wake of global climate change. The rising frequency of extreme weather events has significantly undermined the financial stability of borrowers residing in areas prone to natural hazards, particularly those from underserved and financially vulnerable communities. Several instances have suggested that natural hazards can trigger abnormal volume of mortgagee defaults, potentially caused by destruction of properties and disruption to a family's regular income flow. 
For instance, %\cite{kousky2020flood} investigated a loan-level database that combines post-natural hazard home inspection data, flood zone designations, and loan performance measures in the area impacted by Hurricane Harvey. Among other findings, they 
it was found that loans on moderately to severely damaged homes in areas affected by Hurricane Harvey were more likely to become 90-day delinquent compared to loans on homes with no damage \citep{kousky2020flood}. In the aftermath of Hurricanes Irma and Maria, the home mortgage delinquency rates were tripled in the Houston and Cape Coral metro areas.\footnote{Source: \url{https://realtynewsreport.com/mortgage-delinquencies-tripled-after-hurricane-harvey-foreclosure-threats-follow-all-major-storms-corelogic-study-shows/}} The Tubbs wildfire caused the delinquency rates to spike by 50\% in the Santa Rosa area.\footnote{Source: \url{https://www.corelogic.com/intelligence/wildfires-and-housing-markets/}} %Worse still, residential mortgage defaults only embody a smaller part of the larger crisis, and commercial mortgage defaults constitute another more substantial risk contributor.

We acknowledge the extensive body of economic literature devoted to modeling mortgage defaults \citep[see][for a comprehensive textbook treatment]{leece2008economics}. One strand of the literature focuses on modeling default risk at the individual loan level using classification methods or survival analysis techniques \citep[see][for a comprehensive review and additional references therein]{fitzpatrick2016empirical}. Another stream of literature emphasizes aggregate-level modeling, where regression methods are applied to analyze the percentage of defaults relative to the total number of loans within a given portfolio \citep[e.g.,][]{coleman2005stress,sadhwani2021deep}. A third area of research aims to apply economic models to identify the socio-economic determinants of mortgage defaults \citep[e.g.,][]{elul2010triggers,foote2008negative,foster1984option}.

To the best of our knowledge, no prior statistical study has been devoted for modeling and understanding the subtle relationship between natural hazard losses and mortgage default rates.
This gap is particularly relevant to actuarial considerations, as the co-occurrence of excessive insurance claims due to natural hazards and the accompanying surge in mortgage defaults can create compounded challenges for insurance companies' ALM practices, simultaneously disrupting both the asset and liability sides of their balance sheets. In this paper, we aim to close the gap.  The model developed in this paper, on the one hand, can help verify the significance of this relationship and provides insights into how natural hazard losses translate into increased mortgage default rates. On the other hand, it lays the groundwork for promoting more sophisticated predictive analytics of mortgage default rates in future research.

In light of the discussion above, the goal of this paper is to develop a statistical framework capable of capturing the dependence between natural hazard losses and mortgage default rates. Specifically, we aim to integrate the monthly natural hazard losses extracted from the Spatial Hazard Events and Losses (SHELDUS) database into the evolution models of default rates, calculated based on the Fannie Mae Single-Family Loan Performance (SFLP) data. The details of these two datasets will be provided in Section \ref{sec:data}.  Exploratory analysis of the data suggests that large natural hazard losses may trigger a spike in default rates, but the impact diminishes when the losses are not sufficiently high. 

When formulating a model for the observed time series of default rates and linking it to natural hazard losses, an important consideration is ensuring its interpretability, which enables us to leverage the model estimates to infer whether and how natural hazards influence borrower default behaviors on mortgage loans.
To address the need for both modeling flexibility and interpretability,  we resort to the notion of state space modeling \citep[SSM;][]{durbin2012time}, which have found fruitful applications in a great variety of fields, including actuarial science \citep[e.g.,][]{de1983claims,neves2016forecasting,chukhrova2017state,fung2017unified,youn2023simple,li2024mitigating}. 

SSM provides a unified framework for analyzing how a system evolves over time. A key underlying assumption is that the dynamics of the observed time series, often referred to as space variables, are governed by unobserved state variables. Many popular structural time series models, such as ARMA and ARIMA, can be viewed as special cases of SSM. For our application, we model the observed default rates as our space variable. The time dynamics of these rates are driven by unobserved state variables, which can be broadly interpreted as a summary of the state of the economic system. The effects of natural hazard losses are then assumed to contribute additively through a covariate effect to the (transformed) default rates. This allows us to decouple the effect of the natural hazard losses on default rates from other macroeconomic factors such as interest rates, unemployment rates, housing market conditions, etc., which are summarized by the latent state variables. 

We note that \cite{aktekin2013assessment} also applied SSM to model mortgage default risks.  However, their model's objective was the number of defaults within a given mortgage pool, rather than the default rates, and they did not account for the impact of natural hazard losses. Moreover, in this paper, we deviate from the classical SSM by postulating that natural hazard events with varying levels of losses may have differential impacts on default rates. This is incorporated in the model by considering separate coefficient vectors when losses exceed the threshold and when they do not. 
To estimate the model, we take the Bayesian route, as it provides automatic quantification of uncertainty in parameter estimates, which enables us to draw statistical inference on the hypothesized relationship between natural hazard losses and mortgage default rates. 

{\dr One novel contribution of our paper is to recognize that separating natural hazard losses into normal and extreme regimes is essential for capturing their differential effects on mortgage default rate increases, where the classical SSM fails to address. 
Although the slicing idea may appear conceptually simple at first glance, it is by no means obvious and took considerable efforts for us to recognize. To the best of our knowledge, no prior work has tackled this problem. Our paper thus offers an innovative solution to this modeling gap and provides, at a minimum, a benchmark framework for future research in this direction.
Beyond establishing the statistical significance of natural hazard shocks in triggering surges in mortgage default rates, our model also allows decision-makers to assess the magnitude of these impacts through the estimated regression coefficients. Furthermore, the slicing SSM can serve as a practical tool for predicting adverse changes following natural hazard shocks, which are important quantitative inputs to support decision-making in the potential applications discussed earlier.}

The rest of the article is organized as follows. In Section \ref{sec:data}, we introduce the Fannie Mae SFLP and the SHELDUS datasets. In doing so, we establish plausible relationships between natural hazard losses and mortgage default rates that inform our modeling decisions, in particular the choice of a sliced structure and the selection of relevant months' losses. Next, in Section \ref{sec:model}, we provide an overview of SSM's and introduce the modified sliced variant proposed in this article. A simulation study is included to elucidate how the rarity of natural hazard losses affects regression parameter estimation and provide intuition for the proposed sliced model at hand. In Section \ref{sec:realdata}, 
we demonstrate the fitting of the proposed model to the SFLP and SHELDUS datasets and evaluate its performance using a rolling one-month-ahead prediction procedure, which imitates a real-time use of the proposed model. %{\color{red}We compare the results of the state space model with other standard time series and fixed-time regression models.} 
Section \ref{sec:conclusion} concludes this paper.%, where we also outline future uses of the SFLP and SHELDUS datasets and key areas for improvement in state space modeling for rare events.

\section{Data}
\label{sec:data}

\subsection{Fannie Mae Single-Family Loan Performance Dataset}

%To calculate our response variable, the mortgage default rates within the US, we utilized the Fannie Mae SFLP data. 
The response variable in our study is the mortgage default rate which we source from the Fannie Mae SFLP data. This dataset contains monthly performance data for a subset of, ``30-year and less, fully amortizing, full documentation, single-family, conventional fixed-rate mortgages'' %\citep{fannie_mae_sflp} 
acquired by Fannie Mae since Jan.~1, 2000. We downloaded the data as a ZIP archive containing $96$ quarterly data files comprising Q1 2000 until Q4 2023 through Fannie Mae's Data Dynamics\textsuperscript{\textregistered} portal. We extracted the fields described in Table \ref{table:sflp-fields}, where the definitions of the variables are provided.

\begin{table}
\begin{centering}
    \begin{tabular}{|l|l|p{6.5cm}|}
    \hline
    \textbf{Symbol} & \textbf{Field Name} & \textbf{Description}\\\hline
    \datafield{ID} & Loan Identifier & A unique identifier for the mortgage loan\\\hline
    \datafield{MONTH} & Monthly Reporting Period & The month and year of the loan status information\\\hline
    % \datafield{OD} & Origination Date & \\\hline
    % \datafield{AGE} & Loan Age & \\\hline
    % \datafield{MD} & Maturity Date & \\\hline
    \datafield{STATE} & Property State & A two-letter abbreviation indicating the state or territory within which the property securing
the mortgage loan is located\\\hline
    % \datafield{MSA} & Metropolitan Statistical Area & A numeric code corresponding to the Metropolitan Statistical Area of the property as defined by the US Office of Management and Budget\\\hline
    % \datafield{ZIP} & Zip Code Short & The first three digits of the property's US Postal Service Code \\\hline
    \datafield{STATUS} & Current Loan Delinquency Status & The number of months delinquent as of the reported month\\\hline
    % \datafield{HISTORY} & Loan Payment History & \\\hline
    \end{tabular}
    \caption{Summary of the data fields from the SFLP dataset used in the analysis.} %The descriptions are paraphrased by the authors using information from Fannie Mae's SFLP data dictionary\cite{fannie_mae_data_dict}. }
    \label{table:sflp-fields}
\end{centering}
\end{table}

%\actodo{Add how many data points this data had}
From this data, we can calculate the number of loans in month $t$ that are either current (i.e., up to date on outstanding payments) or are fewer than $k$-months delinquent as

\begin{equation}
C_{k,t} = \sum_{j=1}^{J}\mathbbm{1}_{[0,k)}( \operatorname{STATUS}) *\mathbbm{1}_{\{t\}}(\operatorname{MONTH})\label{eq:dlq-counts}
\end{equation}
% \actodo{I'm guessing in (1) the sum is over j where j runs from 1 to the length of the dataset where each row is a loan. It is better to write it in terms of j instead of rows.}
where each $j$ is one month of data on one loan and $J=2,938,721,940$ is the total number of loan-months in the dataset. %It is worth emphasizing that $C_{k,t}$ includes loans that are either current (i.e., up to date on outstanding payments) or in delinquent statuses.
 Therefore, the percentage of loans that are newly $k$ months delinquent can be computed as
\begin{equation}
    r_{k,t} = \frac{C_{k+1,t} - C_{k,t}}{C_{k+1,t}}\label{eq:dlq-rate}.
\end{equation}
By definition, $C_{k+1, t} \geq C_{k,t}$. %In the above formulation, $r_{k,t}$ represents the proportion of loans that newly entered the $k$ or greater delinquency status in month $t$. 
The denominator in $r_{k,t}$ represents the number of loans that are $k$ or fewer months delinquent. %because if a loan is $k+1$ months delinquent, then it was $k$ months delinquent previously. 
If a loan was previously $k$ months delinquent and the borrower is now up to date on their payments, then the loan will re-enter the denominator. The numerator excludes loans that are below the $(k-1)$-th month delinquency status and beyond the $(k+1)$-{th} month delinquency status, while only keeps the loans \textit{newly} entering into the $k$-months delinquency.  Throughout the rest of this paper, we consider $k=3$ since we treat 90-day delinquencies as mortgage defaults. When there is no risk of confusion, we simplify the default rate notation $r^i_{3,t}$ to $r^i_t$.

In order to calculate $C_{k,t}$ for a specific \datafield{STATE},
% \datafield{ZIP}, or \datafield{MSA},
we add another indicator function to \eqref{eq:dlq-counts} and denote this change by adding a superscript to $C_{k,t}$. Namely, for \datafield{STATE = i}, we write $C_{k,t}^{i}$ where $i \in \{\text{AL}, \ldots, \text{WY}\}$ --- the abbreviations of the variable \texttt{STATE} arranged in alphabetical order. This new state-specific $C_{k,t}^{i}$ is therefore

\begin{equation}
C_{k,t}^{i} = \sum_{j=1}^J \mathbbm{1}_{[0,k)}( \operatorname{STATUS}) *\mathbbm{1}_{\{t\}}(\operatorname{MONTH}) * \mathbbm{1}_{\{i\}} (\operatorname{STATE}).
\end{equation}
Additionally, $r_{t,k}^{i}$ uses the same definition as in \eqref{eq:dlq-rate} with $C_{k,t}$ and $C_{k+1,t}$ replaced by $C_{k,t}^{i}$ and $C_{k+1,t}^{i}$, respectively. 

As an illustration, the left, bottom half of \autoref{fig:texas-loss-dlq} shows the proportion of loans entering 90-day delinquency status, $r_{t}^{i}$, for the state of Texas during the time span of our dataset. Increases in mortgage default rates are clearly observed during the 2008 financial crisis. %The spike in early 2018 corresponds to the aftermath of Hurricane Harvey. %Overall, notice that there appear to be shared factors influencing the mortgage default rate patterns between states, evidenced by the similar trends coinciding with macroeconomic events in the US.
{This lends itself naturally toward a SSM approach, where the involved latent factors can help capture the effects of the changing economic system, allowing us to isolate and analyze the relationship between natural hazard losses and mortgage default rates, which is the focus of this paper.   }
%are allowed to fluctuate independent of our predictors. 

Next, we will explore whether the losses from natural hazards, such as major hurricanes, may have potential explanation capacity 
on the observed mortgage default rates that extend beyond the latent macroeconomic effects. %Throughout this paper, we will consider $k=3$ refer to $r_{3,t}^{i}$, our response variable, as $\rho_t^{i}$. 
 
\subsection{Spatial Hazard Events and Losses Dataset}

\begin{figure}
\begin{centering}

\includegraphics[width=0.55\textwidth]{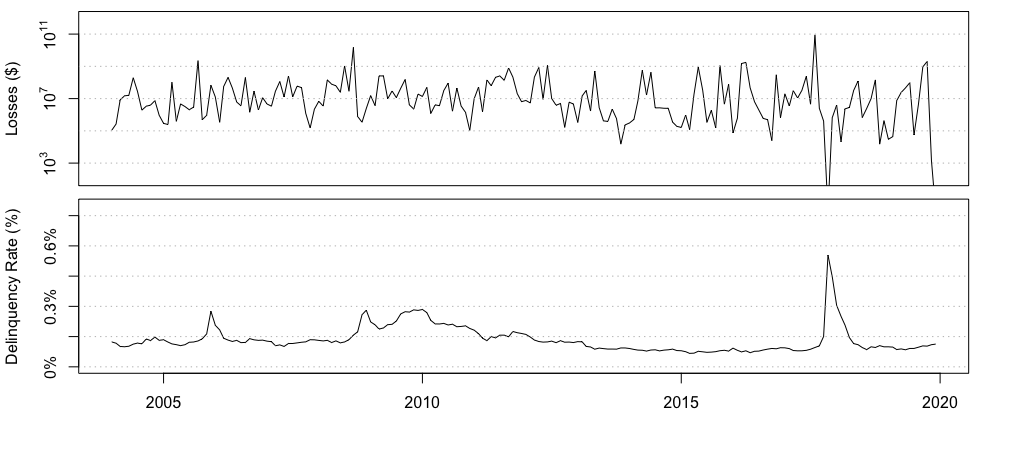}
\includegraphics[width=0.44\textwidth]{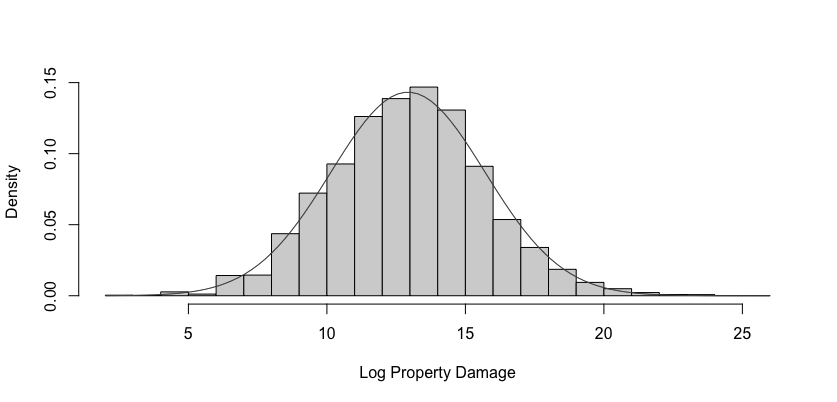}
    
\caption{On the left, monthly natural hazard losses in dollars, $L_t^i$, within the SHELDUS dataset (top) and monthly default rates, $r_{t}^i$, in the Fannie Mae SFLP dataset for Texas. On the right, monthly natural hazard losses at the state level across all states are approximately distributed log-normal with mean $12.927$ and  variance $7.755$.}
\label{fig:texas-loss-dlq}
\end{centering}
\end{figure}

The SHELDUS dataset, % \cite{sheldus21}, 
managed by the Center for Emergency Management and Homeland Security at Arizona State University \footnote{available at \href{https://cemhs.asu.edu/sheldus}{https://cemhs.asu.edu/sheldus}}, enumerates losses due to natural hazards in the US at the county-level. The SHELDUS data includes various forms of direct losses, location, type of natural hazard, and the date of each natural hazard. The losses are inflation-adjusted. To retrieve monthly aggregate damage in each state, we sum the \datafield{PropertyDmg} field grouping by \datafield{Month}, \datafield{Year}, and \datafield{State}. The monthly damages in Texas are shown in the left, top half of \autoref{fig:texas-loss-dlq}.
% \actodo{need to do something about the figures (Fig 1 and) - these are taking too much space without having two pages worth of content }
We index the months and years in a single variable $t$, with $1\leq t \leq T$ and $t \in \mathbb{N}$, and we denote the aggregate loss at time $t$ in state $i$ by $L_{t}^{i}$. 
As shown on the right side of \autoref{fig:texas-loss-dlq}, the state-level monthly losses are approximately distributed log-normal with location parameter  $ \mu_x = 12.927$ and dispersion parameter $ \sigma^2_x=7.755 $.  Moreover, the correlation between losses in subsequent months is $0.023$, indicating that these monthly losses can be treated as uncorrelated (at least linearly). 

To model the relationship between mortgage default rate $r^{i}_{t}$ and natural hazard losses, we introduce a notation to represent the losses in months $t-m$ to $t-n$, inclusive, for state $i$. Because at time $t$, the losses for month $t$ are not yet known, we can only model using the natural hazard losses up to time $t-1$,  thus $n\geq 1$. We denote 
\begin{equation}
    L_{t,m:n}^{i} = \left[L_{t-m}^{i} \quad L_{t-m+1}^{i} \quad \ldots \quad L_{t-n-1}^{i} \quad L_{t-n}^{i} \right]^\top \quad \in \mathbb{R}^{m-n+1}.
\end{equation} 

%Next, we explore the relationship between hazard losses and mortgage default rates. 
As one of the objectives in our modeling, we are interested in knowing which previous month(s) have the most significant impact on the new 3-month delinquency. For instance, it could be that losses from natural hazards take time to manifest as increased mortgage delinquency, in which case the most impactful months of losses for default rates at time $t$ would be before month $t-3$, such as $t-4$ or $t-5$. Alternatively, if effects are immediate, then we would expect month $t-3$ to have the largest impact. %Additionally, we would like to know the nature of the relationship between $X_{t, m:n}^{i}$ and the state-level default rate, $y_t^{i}$. 

\begin{figure}
    \includegraphics[width=\textwidth]{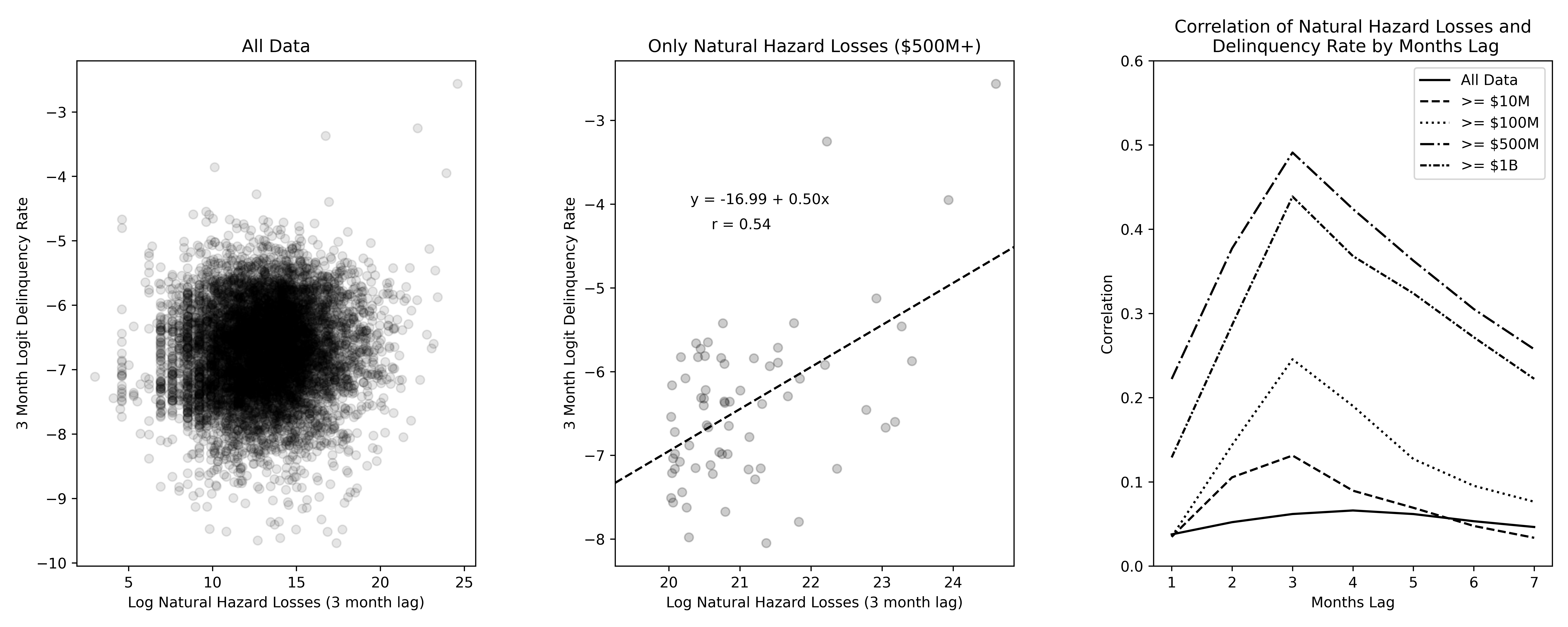}
    \caption{Left panel: The scatter plot between monthly natural hazard losses and the percentage of loans newly entering into 3-month delinquency. Middle panel: The scatter plot between monthly natural hazard losses in log scale and the percentage of loans newly entering into 3-month delinquency for months with natural hazard losses greater than \$$500$ million. Right panel: The correlations between mortgage default rates and natural hazard losses truncated at varying levels, with different lag months.}  
    \label{fig:outlier_loss_dlq}
\end{figure}

The scatter plots and correlation plot in 
\autoref{fig:outlier_loss_dlq} reveal some preliminary insights into the subtle relationship between monthly natural hazard losses and mortgage default rates. Specifically, the scatter plot in the left panel of \autoref{fig:outlier_loss_dlq} shows no clear dependence pattern between the monthly natural hazard losses and mortgage default rates. However, as we shift our focus to months with natural hazard losses exceeding \$500 million, as shown in the middle panel, a positive dependence pattern emerges. The tail dependence between natural hazard losses and mortgage default rates is further apparent in the right panel of \autoref{fig:outlier_loss_dlq}, which summarizes their lagged correlations at different loss truncation levels. Collectively, Figure \ref{fig:outlier_loss_dlq} illustrates that small natural hazard losses appear to have minimal impact on mortgage default rates, while large losses tend to trigger significant increases in default behaviors. This aligns with intuition, as only severe natural hazard events are likely to cause widespread financial hardship within affected communities, thereby impairing individuals' ability to repay their mortgages.
The discussion above suggests that, to uncover the effect of large losses on mortgage default rates, the $L_{t, m:n}^{i}$ values should be segmented based on a threshold informed by the data. 

As for the most influential months, based on the lagged correlations presented in the right panel of \autoref{fig:outlier_loss_dlq}, we hypothesize them to be months $t-3$, $t-4$ and $t-5$. Within our modeling framework, we allow the effects of some of these months to be exactly zero,  and let the data guide the estimation procedure to uncover any supporting evidence. 

% \begin{figure}
% \begin{centering}
    
% \begin{tabular}{|c||c|c|c|c|}
% \hline
% \backslashbox{Lag}{Loss} & \$$10$M$+$ & \$$100$M$+$ & \$$500$M$+$ & \$$1$B$+$ \\\hline\hline
% 1 Month & 0.035 & 0.035 & 0.222 & 0.129 \\\hline
% 2 Months & 0.105 & 0.143 & 0.377 & 0.287 \\\hline
% %\rowcolor{green!20}
% 3 Months & 0.131 & 0.244 & 0.490 & 0.438 \\\hline
% 4 Months & 0.089 & 0.189 & 0.423 & 0.367 \\\hline
% 5 Months & 0.069 & 0.127 & 0.362 & 0.323 \\\hline
% 6 Months & 0.048 & 0.095 & 0.305 & 0.271 \\\hline
% 7 Months & 0.034 & 0.076 & 0.257 & 0.222  \\\hline
% \end{tabular}
% \caption{Correlation between log losses and log delinquency rates at the state level for various choices of lag and loss cutoffs. A lag of 3 months consistently has the highest correlation.}
% \label{table:important-month}

% \end{centering}
% \end{figure}

\section{State Space Modeling}
\label{sec:model}

\subsection{Overview of Classical State Space Modeling}\label{sec:state_space_model}

SSM is a hierarchical data modeling framework consisting of a set of %(potentially dependent)
latent stochastic processes that collectively govern an observed variable.   %Our inferential goal is estimation of the latent states and model parameters to enable prediction of future mortgage default rates. 
In this paper, we are interested in a class of SSM's that integrate covariate information through a linear model in the observation space and incorporate a local linear trend structure in the state space. That is
\begin{equation}
    \begin{aligned}
y_t &=X_t^\top \beta+\mu_t+\epsilon_t,  &&  \epsilon_t \sim N\left(0, \sigma_y^2\right) \\
\alpha_t = \begin{pmatrix}\mu_t \\ \nu_t\end{pmatrix} &= A \alpha_{t-1}+ \eta_t,  && \eta_t \sim N_2(0, Q), \,\, A = \begin{pmatrix}
    1 & 1\\
    0 & 1
\end{pmatrix},
\label{eq:model}
\end{aligned}
\end{equation}
where $Q = \text{diag}(\sigma^2_\mu, \sigma^2_\nu)$, and $t = 1, \ldots, T$. We also assume $\alpha_0 = \mathbf{0}$. When $\sigma^2_\nu = 0$, the model collapses to a deterministic trend model. Borrowing the notation from the previous section, $y_t$ is a function of the 3-month delinquency rates, $f(r_t^i)$, from the Fannie Mae SFLP dataset and $X_t$ is a function of the losses, $g (L_{t, m:n}^i)$, from the SHELDUS dataset. The functions $f$ and $g$ correspond to appropriate data transformations, which will be specified in Section \ref{sec:model-delinquency}. {\dr The state components, $\alpha_t = (\mu_t \quad \nu_t)^\top$, reflect a linear trend with a random walk slope, $\nu_t$, that accumulates with local Gaussian perturbations into $\mu_t$.  These components are designed to capture underlying dynamics that cannot be explained by the observed covariates, thereby allowing us to isolate the effects of natural hazard loss variables on mortgage default rates. In addition to the impact of natural disaster losses, another important driver of changes in mortgage default rates is the broader macroeconomic environment, which we interpret as being captured by the latent state variables. The dynamic structure we assume for the latent factor for representing the evolution of the macroeconomic environment, is one of the standard forms commonly adopted in economics-related state-space modeling problems \citep[e.g.,][]{abdallah2013evidence,chan2023bayesian}.
}

{\dr It is noteworthy that in the proposed modeling framework, the macroeconomic environment is not treated as an exogenously specified factor. Instead, it is captured endogenously through the latent state variable $\mu_t$, which is estimated using Bayesian methods. The rationale for incorporating the macroeconomic environment implicitly as a latent variable is twofold. First, due to the complexity of macroeconomic dynamics and their effects on mortgage default rates, it is challenging to identify all relevant macroeconomic factors in order to isolate the effects of natural hazard shocks. %or to construct a single summary statistic that fully captures their impact. 
Second, even if all significant macroeconomic determinants could be identified, predicting mortgage default rates would require first developing predictive models for each of these factors, adding another layer of complexity to the modeling task. Using a latent factor to represent the macroeconomic evolution affecting mortgage default rates addresses both of these challenges, enabling us to focus on examining the relationship between natural hazard losses and mortgage default rates, which is the primary objective of this paper.}
%These components are designed to capture the fluctuations arising from other macroeconomic factors, enabling us to isolate the effects of natural hazard loss covariates on mortgage default rates.

%The main benefit of using a SSM is the persistence of latent factors over time to capture other macroeconomic factors driving changes in mortgage default rates. Because the aim of our study is to determine the relationship between mortgage default rates and hazard losses, accounting for these macroeconomic factors allows us to isolate the effects of the covariates on mortgage delinquency rates %since SSMs are a popular tool to study macroeconomic behavior over time; see for example \cite{chan2023bayesian}.
%Additionally, regarding 

{\dr Concerning the interpretability of the regression parameters in the proposed model, let us remark that 
$y_t$ is {conditionally} independent of $\mu_{t-1}$ given $\mu_t$, but unconditionally, they are dependent. Roughly, this setting treats $\mu_t$ as a summary of all the past dynamics up to time $t$, carrying forward the effects of past shocks, trends, and any persistent patterns.
Since $\mu_t$ already incorporate all relevant information from $\mu_0,\mu_1,\ldots,\mu_{t-1}$, $y_t$ becomes conditionally independent of $\mu_{t-1}$ once $\mu_t$ is known. This is the classical first-order Markov assumption commonly imposed on the latent state in SSM. Consequently, given $\mu_t$, the regression coefficient $\beta$ associated with the hazard loss variable $X_t$ can directly reflect the extent of the linear relationship between $X_t$ and $y_t-\mu_t$, without the need to account separately for the influence of earlier latent dynamics. This provides a clear and interpretable meaning for the estimate of $\beta$.
}
%the observation space, $y_t$, is independent of the latent state at time $t-1$, $\mu_{t-1}$, conditional on $\mu_t$, meaning that $\beta$ directly describes the linear relationship between $X_t$ and $y_t - \mu_t$. }

Let $\theta = (\beta$, $\sigma_y^2$, $\sigma^2_\mu, \sigma^2_\nu)$ denote the vector of parameters associated with the SSM specified in Equation \eqref{eq:model}. While a classical frequentist approach can be adopted to estimate $\theta$, we take the Bayesian route due to its natural ability to quantify uncertainty, thereby facilitating inference on the significance of the covariate effects. To this end, we endow $\theta$ with a prior distribution. Specifically, we use a product prior: $\pi(\theta) \propto \pi(\beta) \pi(\sigma_y^2)  \pi(\sigma^2_\mu)  \pi(\sigma^2_\nu)$. Then by Bayes theorem, the posterior distribution of $\theta\mid (y_t, X_t)_{t=1}^T$ is given by 
$$\pi(\theta \mid (y_t, X_t)_{t=1}^T) \propto p(y_1, \ldots, y_T\mid X_1, \ldots, X_T, \theta) \pi(\theta).$$

Markov Chain Monte Carlo sampling can be used to sample from this posterior to compute posterior summaries of parameters. However, a key barrier in fitting the model \eqref{eq:model} lies in handling the latent components $\alpha_t = (\mu_t \quad \nu_t)^\top$. Note that the likelihood of the observed data is given by  $$p(y_1, \ldots,y_T\mid  X_1, \ldots, X_T, \theta) = \int \left[\prod_{t=1}^T p(y_t \mid \alpha_t, \theta)\right] p(\alpha_t) d\alpha_t. $$ Hence, in order to sample from this posterior distribution, one needs to deal with the high-dimensional integral involved in the observed likelihood. This issue can be addressed by what is known as the data-augmentation trick \citep{hobert2011data}. The idea is to augment the target posterior distribution by incorporating the conditional distribution of the latent states given the observed data:
\begin{align*}
    \pi(\theta, (\alpha_t)_{t=1}^T \mid (y_t, X_t)_{t=1}^T) \propto \left[\prod_{t=1}^T p(y_t \mid \alpha_t, \theta)\right] p(\alpha_t) \pi(\theta).
\end{align*}
A standard Gibbs sampler would then be applied to sample $\theta \mid (\alpha_t, y_t, X_t)_{t=1}^T $ and $\alpha_t \mid \theta, (y_t, X_t)_{t=1}^T$. %Suppose for now we have a way of sampling $\alpha_t \mid \theta, (y_t, X_t)_{t=1}^T$. 
The advantage of adopting the augmentation approach is that, conditional on the latent variables $(\mu_t, \alpha_t)$, the model in \eqref{eq:model} simplifies to a basic linear regression model. Moreover, the approach allows prior specification for $\theta \mid \alpha_t$ to be straightforward and intuitive. For example, in our work, we use a spike and slab prior \citep{george1993variable} on the regression coefficients $\beta$, namely, %, which helps facilitate variable selection by distinguishing between significant and negligible covariate effects. Namely, we let %for $j = 1, \ldots, p$,
$$\beta_j\mid (\alpha_t)_{t=1}^T \overset{\text{i.i.d}}{\sim} \rho \Gauss(0, \tau^2) + (1-\rho) \delta_0,$$ 
where $\delta_0$ is a point mass at $0$. The implication of this prior choice is that with probability $\rho$, the variable $\beta_j \mid (\alpha_t)_{t=1}^T \sim \Gauss(0, \tau^2)$ and with probability $(1-\rho)$, it is set to 0. The advantage of using this class of priors over other common continuous priors, such as a normal distribution, is that the posterior probability of $\beta_j=0$ is positive, which allows for direct inference about significant or negligible covariate effects. %This is not the case for continuous priors since for them this probability is always exactly equal to zero.  
The parameter $\rho$ encodes our prior belief on the probability that a particular regression component will be included in the model. We set $\rho = 0.5$ in our analysis.

For the variance parameters $\sigma_y^2, \sigma^2_\mu, \sigma^2_\nu$, we assume conditionally conjugate priors, i.e., $\sigma_y^2 \mid (\alpha_t)_{t=1}^T \sim \text{Inverse-Gamma}(a, b)$  with $a> 0$, $b >0$. In our analysis, we set both $a$ and $b$ to small positive numbers, which is a common choice of non-informative hyperparameters. For the sake of completeness, we remind the reader that a random variable $Z$ is said to follow an inverse gamma distribution with parameters $a, b > 0$, denoted by $Z \sim \text{Inverse-Gamma}(a, b)$, if its density $f(z) \propto z^{-a-1}e^{-b/z}$ for $z > 0$.

Although spike and slab priors are suitable for variable selection and interpretation, they do not render easy posterior sampling. This problem is typically addressed by introducing another set of latent variables $\gamma = (\gamma_1, \ldots, \gamma_p)$ such that $\beta_j \mid \gamma_j = 1, (\alpha_t)_{t=1}^T \sim \Gauss(0, \tau^2)$ and $\beta_j \mid \gamma_j = 0, (\alpha_t)_{t=1}^T \sim \delta_0$. Clearly, $\gamma_j \overset{\text{i.i.d}}{\sim} \text{Bernoulli}(\rho)$, and marginalizing over $\gamma_j$ yields the desired prior distribution. Moreover, given $\gamma$, the model \eqref{eq:model} can be expressed as
\begin{align*}
    y_t = X_{t,\gamma}^\top \beta_\gamma + \mu_t + \epsilon_t \\
    \alpha_t = A \alpha_{t-1} + \eta_t, 
\end{align*}
where $X_{t,\gamma}$ is a $|\gamma| \times 1$ vector with columns equal to $X_t$ if and only if the corresponding $\gamma_j = 1$; $\beta_\gamma$ is similarly defined. Let $\phi \mid -$ denote the conditional distribution of any generic parameter $\phi$ given all other parameters and the observed data. Then, given $\gamma$ and $(\alpha_t)_{t=1}^T$, we have the following full conditional distributions:
\begin{align*}
    \beta_\gamma \mid - &\sim \Gauss(\Omega_\gamma X_\gamma^\top (y - \mu), \Omega_\gamma), \quad \Omega_\gamma = (X_\gamma^\top X_\gamma + \tau^2 \mathrm{I}_{|\gamma|})^{-1};\\
    \sigma_y^2 &\mid - \sim \text{Inverse-Gamma}\left( \frac{T}{2} + a, \frac{1}{2} \sum_{t=1}^T (y_t - \mu_t -X_{t,\gamma}^\top \beta )^2\right);\\
    \sigma^2_\mu &\mid - \sim \text{Inverse-Gamma}\left( \frac{T}{2} + a, \frac{1}{2} \sum_{t=1}^T (\mu_t - \mu_{t-1} -\alpha_t )^2\right);\\
    \sigma^2_\nu &\mid - \sim \text{Inverse-Gamma}\left( \frac{T}{2} + a, \frac{1}{2} \sum_{t=1}^T (\alpha_t -\alpha_{t-1} )^2\right).
\end{align*}
The above steps are complemented by the update of the latent indicators $\gamma$ using Bayes theorem:
\begin{align*}
    p(\gamma_j = 1 \mid \gamma_{-j}, -) \propto p(y_1, \ldots, y_T \mid (\alpha_t)_{t=1}^T ,X_{\gamma_{-j}}) \rho,
\end{align*}
where $\gamma_{-j}$ is a vector obtained by deleting the $j$-th element of $\gamma$. Essentially, the first term in the right-hand side of the above expression computes the likelihood of the observed data given that the $j$-th variable is included in the model. This is then multiplied by the prior probability that $\gamma_j = 1$ which is $\rho$. 
Throughout the remainder of this paper, we shall denote this basic model as {\bf SSM}.
%Finally, $\rho$ is updated from its full conditional which is $\text{Beta}(|\gamma|, p - |\gamma|)$.

Next, we come to the sampling of $\alpha_t \mid \theta, (y_t, X_t)_{t=1}^T$. This is done by combining sequential Monte Carlo \citep{doucet2001sequential} and Kalman filtering. We briefly describe the basics. Suppose we are given $\pi(\alpha_{1:t} \mid y_{1:t})$. Then
\begin{align*}
    \pi(\alpha_{1:(t+1)} \mid y_{1:(t+1)})&  = \dfrac{\pi(y_{1:(t+1)} \mid \alpha_{1:(t+1)}) \pi(\alpha_{1:(t+1)})}{\pi(y_{1:(t+1)})} \\
    & = \dfrac{\pi(y_{1:t} \mid \alpha_{1:t}) \pi(y_{t+1}\mid \alpha_{t+1}) \pi(\alpha_{t+1} \mid \alpha_t)\pi(\alpha_{1:t})}{\pi(y_{1:t}) \pi(y_{t+1}\mid y_{1:t})}\\
    & = \dfrac{\pi(\alpha_{1:t} \mid y_{1:t})\pi(\alpha_{t+1} \mid \alpha_t) \pi(y_{t+1}\mid \alpha_{t+1}) }{\pi(y_{t+1}\mid y_{1:t})},
\end{align*}
where the reduction in the conditioned event holds due to the Markov nature of the model. We note here that conditioning on $\theta$ is implicit. In the above expression, the distributions $\pi(\alpha_{t+1} \mid \alpha_t)$ and $\pi(y_{t+1}\mid \alpha_{t+1})$ are completely known. A key thing to note here is that the denominator is independent of $\alpha_{1:(t+1)}$. The required expressions are then computed using the Kalman filter \citep{durbin2012time}.

%In the model described by \autoref{eq:model}, we needed to fit $\beta$, $\alpha_{0:T}$, $\sigma^2$, $\eta^2_\mu$, and $\eta^2_\nu$. We used the Bayesian Structure Time Series (BSTS) \cite{scott_bsts_2024} package in the R programming language to generate Monte Carlo samples from the posterior distribution for each of these parameters given $Y_{1:T}$. In general, we used 5000 samples with the first 2000 samples discarded as burn-in. Using these posterior samples, we predicted mortgage default rates one time-step ahead, which mocks a real-time, regularly updated fitting and prediction procedure. When determining the predictive error of the model, we primarily considered the model's early detection capabilities. That is, if there is a spike in the covariate values that preceded a spike in the observation level, how early could that observation-level spike be detected?

\subsection{The Modified Sliced State Space Modeling}
The preliminary analysis presented in the right panel of Figure \ref{fig:outlier_loss_dlq} indicates that the significance of the impact of natural hazard losses on mortgage default rates depends on the magnitude of the losses. The direction of dependence is also quite evident in that natural hazard losses below a certain threshold have minimal impact on default rates, whereas higher losses exert a significantly stronger influence. The classical SSM specified in \eqref{eq:model} does not account for such varying effects of the covariates. We now propose a modified sliced SSM which incorporates additional flexibility that allows us to capture the subtle relationship between natural hazard losses and mortgage default rates, which emerges only in the extreme scenarios. Then we argue that the proposed modified model can be can be reformulated into a structure consistent with the one in \eqref{eq:model}, and thus similar estimation procedure can be used to implement the proposed modified model. %{\color{red}Consider the following model:
%\begin{align}\label{eq:modified_model}
%  \nonumber  y_t &= \begin{cases}
%        X_t^\top \beta^l + \mu_t + \epsilon_t \quad \text{if } X_t \leq c \\
%        X_t^\top \beta^u + \mu_t + \epsilon_t \quad \text{if } X_t > c 
%    \end{cases} \\
%    \alpha_t &= A\alpha_{t-1} + \eta_t, 
%\end{align}}

Consider the following model:
\begin{align}\label{eq:modified_model}
  \nonumber  y_t &=
        (X_t^{l})^\top \beta^l + (X_t^{u})^\top \beta^u + \mu_t + \epsilon_t
    \\
    \alpha_t &= A\alpha_{t-1} + \eta_t, 
\end{align}

\noindent where $X_t^{u}$ are the values of $X_t$ that exceed the threshold with zeros elsewhere, $X_t^{l}$ are the values of $X_t$ at or below the threshold with zeros elsewhere, $c$ is the threshold, and $\alpha_t$ is defined as in \eqref{eq:model}. The parameter $\beta^l$ in \eqref{eq:modified_model} captures the effect of natural hazard losses on mortgage default rates that do not exceed the threshold $c$. On the other hand, $\beta^u$ is the effect when the losses exceed the threshold $c$. From our preliminary analysis, we expect $\beta^l$ to be small, which means when losses are small, mortgage default rates are mostly governed by the economic attributes captured by the latent components $\alpha_t = (\mu_t \quad \nu_t)^\top$.  %However, when the losses are small, the impact of them is captured by $\beta^u$. 
The introduction of separate coefficients for the two cases does not mean that we consider two separate processes of default rates since these are connected by the same underlying latent state $\alpha_t$.
The model is a special case of a varying coefficients model, where one assumes $y_t = X_t^\top \beta_t + \mu_t + \epsilon_t$, and models $\beta_t$ as a function of $t$. In our case, $\beta_t$ has two distinct values $\beta^l$, $\beta^u$.  Compared with the more general approach with varying coefficients, the proposed sliced SSM offers the advantages of easier implementation and more straightforward interpretation.  

Bayesian inference on this modified model can be recast to the posterior sampling procedure described in Section \ref{sec:state_space_model}. %Indeed, define the variable $I_t = 1$ if any of the elements $X_t$ exceeds the threshold $c$ and set $I_t = 0$ otherwise. Next, 
Specifically, define a new covariate vector $\widetilde{X}_t = (X_t \mathbb{I}_{X_t\geq c}, X_t \mathbb{I}_{X_t< c})$, and let $\tilde{\beta} = (\beta^l, \beta^u)$. 
Then, the modified model \eqref{eq:modified_model} can be reformulated as
\begin{align*}
    y_t = \widetilde{X}_t \tilde{\beta} + \mu_t + \epsilon_t \\
    \alpha_t = A\alpha_{t-1} + \eta_t,
    \end{align*}
which is consistent with the structure considered in \eqref{eq:model}. The prior choices for other parameters remain the same as discussed in the previous subsection. Moreover, we assign $\tilde{\beta}_j\mid (\alpha_t)_{t=1}^T \overset{\text{i.i.d}}{\sim} \rho \Gauss(0, \tau^2) + (1-\rho)\delta_0$.
Throughout the rest of this paper, we shall denote this modified model as {\bf mSSM}.
%We also consider a hierarchical version of this model wherein we conduct a joint analysis of all the states together. This is discussed in the Appendix.

%From a computational point of view also, this hierarchical model is appealing in the sense that all updates discussed above for the non-hierarchical model would carry over into this model due to the conditional independence of $\beta^{u,i}$ given $\beta_0$.

{\dr In concluding the discussion of our proposed methodology, we remark that although all the SSM's we have outlined so far contain only linear structures, as we will see in a moment in the real data analysis in Section \ref{sec:model-delinquency}, we set $X_t$ to be the standardized logarithmic transformation of the natural hazard losses, and we apply the logit transformation to the mortgage default rate as the response variable.  These transformations already introduce a nonlinear relationship between natural hazard shocks and mortgage default rates. We have experimented with other transformations of both the explanatory and response variables, and the selected forms resulted in the best model fit to the data. In this sense, we have effectively performed a manual tuning of the nonlinear functional form in the model.

Our proposed modeling framework could, in principle, incorporate more flexible semi-parametric or non-parametric regression structures such as splines, which would allow for a more automated specification of the nonlinear relationship in the model. However, such approaches would substantially weaken interpretability, which is a key consideration in this paper, as our primary objective is to identify statistical evidence that natural hazard shocks can trigger surges in mortgage default rates. In addition, they may introduce robustness concerns for our slicing method, particularly given the limited amount of training data available for estimating the regression coefficient in the high-loss regime. For these reasons, we focus in this paper on a parametric SSM rather than incorporating nonparametric methods.
}

\subsection{A Simulation Study}
\label{sec:sim}

\begin{figure}[h]
\begin{centering}
    \includegraphics[width=0.85\textwidth]{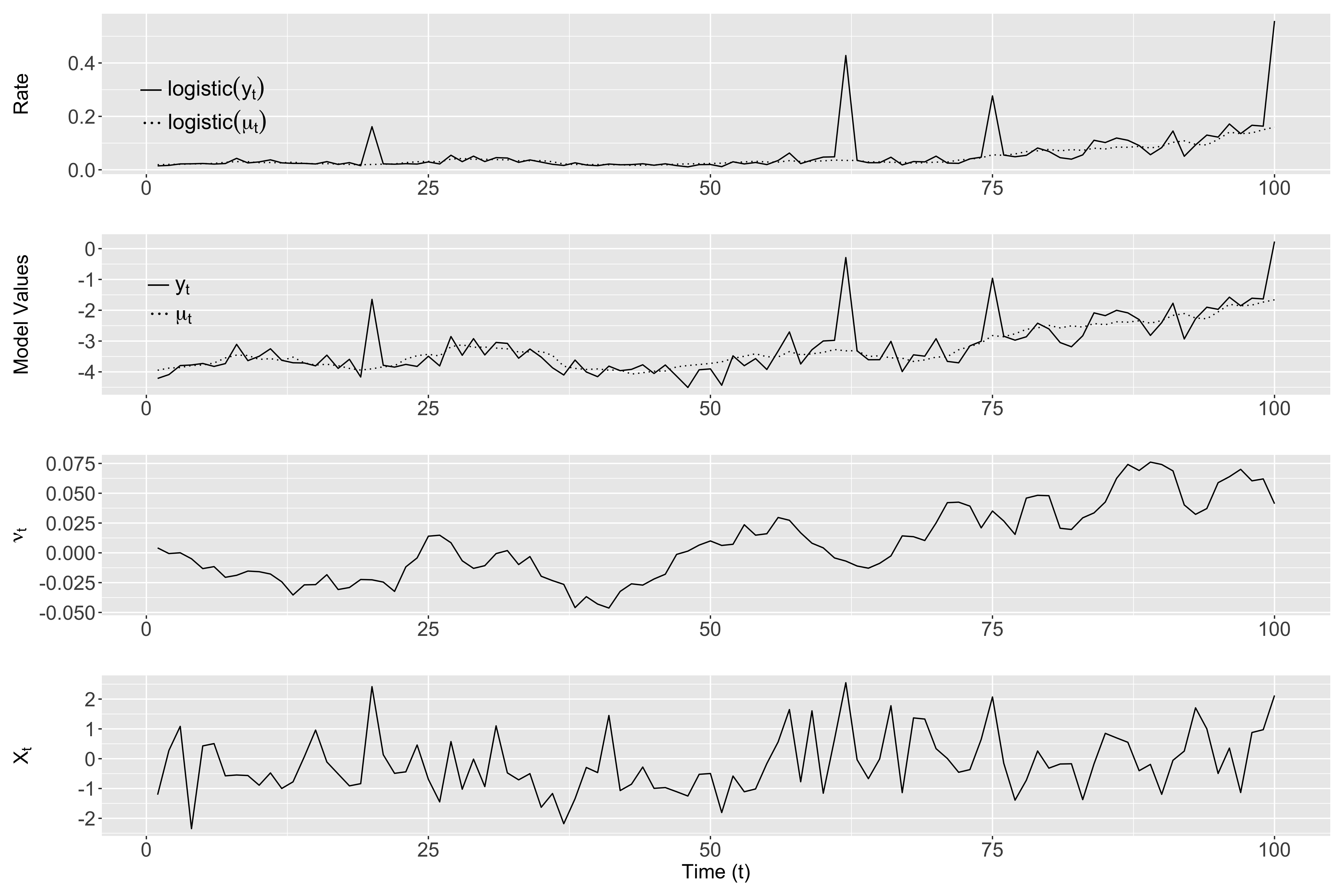}
    \caption{Illustration of a realization of the simulated observed state, latent state, and predictor, according to the data generation process described at the beginning of Section \ref{sec:sim}. }
    %A simulation of the local linear trend model with $\beta^u = 10$, $\beta^l = -0.5$, $\alpha_0 = (-6 \quad 0)^\top$, $\sigma_y^2 = 0.25$, $\sigma_\mu^2 = 10^{-2}$, $\sigma_\nu^2 = 10^{-4}$, and $ X_t \overset{\text{i.i.d}}{\sim} \Gauss(0, 1)$.}
    \label{fig:simple_llt}
\end{centering}
\end{figure}

To elucidate the behavior of modified sliced SSM defined in Equation \eqref{eq:modified_model} as well as its effectiveness for capturing the tail dependence in disaster losses and mortgage default rates, we simulated the process with $\beta^u = 1$, $\beta^l = 0.1$, $\alpha_0 = (-4 \quad 0)^\top$, $\sigma_y^2 = 0.25$, $\sigma_\mu^2 = 10^{-2}$, $\sigma_\nu^2 = 10^{-2}$, and $X_t \overset{\text{i.i.d}}{\sim} \mathrm{N}(0, 1)$.  Moreover, we set the slicing threshold to $c=2$.  \autoref{fig:simple_llt} shows a realization of this process, where we can see spikes in $ X_t$ correspond to spikes in $y_t$. We present the values of $\operatorname{logistic}(y_t)$ and $\operatorname{logistic}(\mu_t)$ in the top chart of \autoref{fig:simple_llt}, which transforms the observation state to a rate on the interval $(0,1)$, similarly to our observed mortgage default rate data. 

{\dr Furthermore, the values of $y_t$ approximately center around the state space value, with short-term perturbations due to $X_t$. Over time, $y_t$ may increase or decrease substantially due to $\mu_t$, suggesting that controlling for state-space variation is important for isolating the effects of $X_t$ on $y_t$. During time periods without natural hazard shocks,  the distribution of the transformed observed mortgage default rates is roughly $\mathrm{N}(\mu_t, \sigma^2_y)$. Hence, given $\mu_t$, the variation of $y_t$ would then lie within the range, says, $[\mu_t - 3\sigma_y, \mu_t + 3\sigma_y]$ with high probability. Deviations beyond this range are highly improbable under normal conditions, and they are likely indicative of exogenous shock effects. It is precisely this type of unusually large, shock-induced variation that constitutes the primary focus of our study.}

Next, we assume that the parameter values used in the data-generating process are unknown, and we evaluate the performance of the rolling window estimation of the regression coefficients $\beta^u$ and $\beta^l$ using the Bayesian method described in Section \ref{sec:model}, along with the associated prediction performance of the estimated model. {\dr By rolling-window, we refer to the process in which the time point up to which data is available incrementally advances, while the underlying model specification remains unchanged. Specifically, let $\hat{\beta}^{u,T}$ and $\hat{\beta}^{l,T}$ denote the estimated regression coefficients based on data from time period $1:T$. When new data becomes available for period $(T+1)$, these estimates are updated by re-fitting the same model to the extended dataset covering $1:(T+1)$.}

\begin{figure}
\begin{centering}
    \includegraphics[width=0.75\textwidth]{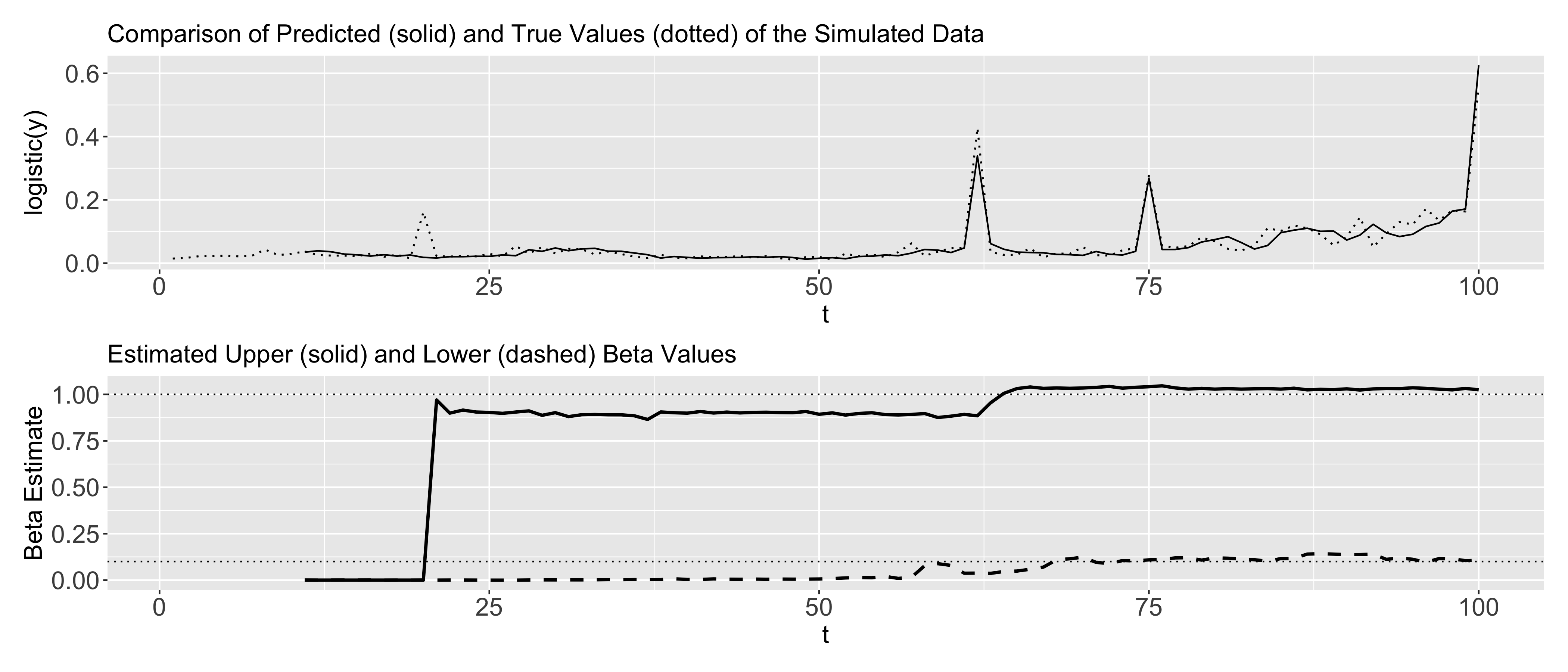}
    \caption{Illustration of the estimation performance of the proposed Bayesian method and the prediction performance of the estimated model. The top panel presents one-step posterior mean predictions, while the bottom panel displays the estimated values of $\beta^l$ and $\beta^u$.}
    \label{fig:thresh-beta}
\end{centering}
\end{figure}

The numerical results are summarized in Figure \ref{fig:thresh-beta}.
Several key findings emerge. First, the estimate of $\beta^u$ remains near zero until around time $t = 20$, when the first natural hazard shock occurs. As a result, the model fails to anticipate the initial spike in mortgage default rates, which is natural to expect because the model has not yet encountered a similar shock in its training window. However, once the first shock is observed, the estimation procedure rapidly detects the signal, and the estimated value of $\beta^u$ quickly adjusts toward its true value. From that point onward, the model is able to accurately capture subsequent shock-induced jumps in mortgage defaults.
Meanwhile, the estimate of $\beta^l$ remains closely aligned with its true value throughout the entire period considered. Overall, these results demonstrate that the proposed estimation procedure efficiently recovers the structural effects of rare natural hazard shocks on mortgage default risk.

%Next, we explore the impact of a rolling window covariate structure, with the covariate data split at a threshold $c$ using a local level model. Based on \autoref{fig:outlier_loss_dlq}, the linear relationship between $\log L_{t,m:n}$ and $\log y_t$ becomes stronger as the cutoff of losses increases, with a maximum correlation when considering only losses greater than \$500 million. We define two vectors $ X_t \mathbb{I}_{ X_t\geq c}$ and $ X_t \mathbb{I}_{X_t< c}$ with corresponding regression parameters $\beta^{u}$ and $\beta^{l}$ for the upper and lower covariate values. The key takeaway from modeling in this manner is that if there have been no events with catastrophic level losses, then the fitting procedure will not be able to learn $\beta$. However, if there is even just one event, these estimates become stable, as shown in \autoref{fig:thresh-beta}. 

\section{Real Data Analysis}
\label{sec:realdata}

In order to implement the model in \eqref{eq:modified_model} on the Fannie Mae SFLP and SHELDUS datasets, there are several issues we need to consider, including selecting appropriate data transformations, determining the number of months of covariates to include, and deciding whether to bifurcate the natural hazard losses at a dollar amount threshold for the regression analysis.  These considerations are addressed in the succeeding subsections.  

\subsection{Data Transformations}
\label{sec:model-delinquency}

We begin our analysis by considering a reasonable transformation for the explanatory variable, $L_t^i$. We aim to find a function $g(x): \mathbb{R} \rightarrow \mathbb{R}$ such that the empirical distribution of $g(x)$ has few outliers. %and can be utilized across multiple states. 
As shown in \autoref{fig:texas-loss-dlq}, the distribution of natural hazard losses is approximately log-normal. This bodes well for estimation of regression parameters, because under a $\log$ transform, it is unlikely that we will experience extreme values of the covariates that could skew the estimate of $\beta$. %Additionally, we have that the autocorrelation of lag 1 for these variables is near zero, which indicates near-decorrelation of month-over-month losses. This is convenient for the explainability of our model, because all of the retained information from $t$ to $t+1$ is assumed to live in the latent states. Also, 
To further improve the interpretation of the regression coefficients, we scale the covariates via 
$$g(x) = \frac{\log x - \mu_x}{\sigma_x}$$
where $\mu_x = 12.927$ and $\sigma_x = \sqrt{7.755}$ %are normalization constants as estimated across all states collectively and 
as demonstrated in \autoref{fig:texas-loss-dlq}.

\begin{figure}
    \begin{centering}
    \includegraphics[width=\textwidth, height = 6cm]{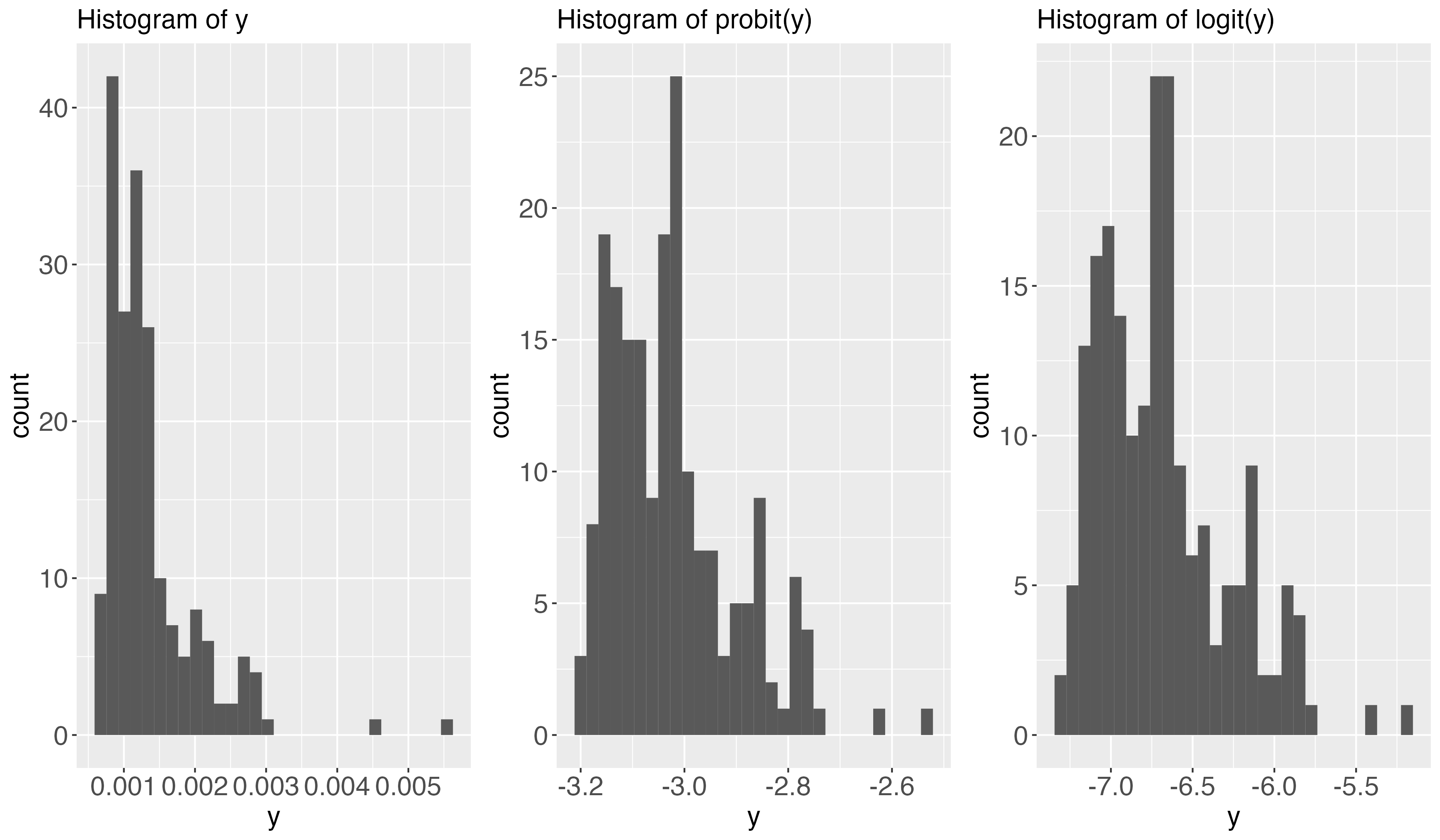}
    \caption{Change in the shape of the distribution for 90-day mortgage delinquency rates in Texas resulting from logit and probit transformations.}
    \label{fig:probit_logit_hist}
    \end{centering}
\end{figure}

Similarly, for the response variable, $\rho_t^i$, we need to identify an invertible function $h: [0,1] \rightarrow \mathbb{R}$ that ensures the residuals from fitting \eqref{eq:modified_model} are approximately normal distributed. Two natural options are the logistic and probit transformations given as $$\operatorname{logit}(y) = \log\frac{y}{1-y} \quad \text{and} \quad \operatorname{probit}(y) = \Phi^{-1}(y)$$
where $\Phi^{-1}(y)$ is the inverse standard Gaussian CDF. Both of these transformations produce reasonably similar results when applied to our dataset, as shown in \autoref{fig:probit_logit_hist}. Because the logit is a mathematically simpler transformation, we will default to its use throughout this analysis. 

%Next, we consider whether the logit transformed data produces reasonable residuals directly or if any further rescaling should be applied. A common approach for finding an appropriate transformation is to try fitting the model with a Box-Cox transformation applied to the response with different values of $\lambda$. In our case, because we have already transformed the data with the logit, our full transformation becomes
%$$f(y) = \begin{cases}
%    \begin{aligned}\frac{(-1*\operatorname{logit}y)^\lambda + 1}{\lambda} && \text{if } \lambda \neq 0\\
%    \log( -1*\operatorname{logit}y) && \text{if }\lambda = 0.
%    \end{aligned}
%\end{cases}$$
%The $-1$ in front of the logit transformations comes from the need for the input to a Box-Cox transformation to be positive. If fed negative values, some outputs will become complex numbers. This change means that what we are actually modeling is a Box-Cox transformation of the mortgage \textit{non}-delinquency rates. As for choosing an appropriate value for $\lambda$, we fit the model using $g(X_{t,5:3}^{\text{Texas}})$ as our explanatory variable and $y_t^{\text{Texas}}$ as our response for $\lambda \in \{-2, -1.5, -1,\dots, 1.5, 2\} $ and found that $\lambda =1$, corresponding to the identity function, is reasonable to use, and we choose it out of parsimony. This simple result also means that the $-1$ multiplied against the logit is unnecessary, and we return to modeling delinquency rates directly.

%\begin{figure}
%    \begin{centering}

%    \includegraphics[width=\textwidth]{graphics/boxcox_qq.png}
%    \caption{Various values of $\lambda$ from a Box-Cox transformation on $y_t^i$ for Texas when fit with using $X_{t,5:3}^{\text{Texas}}$ as covariates.}
%    \label{fig:boxcox_qq}

%    \end{centering}
%\end{figure}

\subsection{Variable Selection and Slicing}

In our model, we have to decide which months of losses to include as covariates at time $t$. Our analysis in \autoref{fig:outlier_loss_dlq} indicates that for $3$ month delinquency rates, the losses in month $t-3$ have the highest correlation with the delinquency rate. %The high correlation for month 3 holds across various loss levels, ranging from all losses to only losses greater than $\$1$ billion.
Furthermore, because losses in months $t-1$ and $t-2$ occurred less than three months prior to the observed 3-month delinquency rates, they are intentionally excluded from our model. 

\autoref{fig:outlier_loss_dlq} also demonstrates that at low levels of losses from natural hazard, there appears to be no significant relationship between the losses and mortgage default rates. This comes as no surprise, as less severe natural hazard events are less likely to impose significant financial hardships on borrowers, or may not result in a sufficient number of severely damaged properties to cause a noticeable shift in default rates across a large geographic area, such as a state. However, once losses reach a certain substantial level, a significant relationship emerges between natural hazard losses and default rates.
To incorporate knowledge of this multifaceted relationship between the covariates and our response, we split the covariates into two regimes, one where losses exceed $\$10$ billion, and another where losses are under $\$10$ billion.%, denoted by $X$ in \autoref{eq:modified_model}. 

{\dr We openly admit that the choice of the \$10 billion threshold involves a degree of subjectivity. In the data, we observed that natural hazards with losses exceeding \$10 billion appeared to be less effectively absorbed by the state economies, and mortgage default rate changes following such extreme events were approximately three times larger than those following smaller losses in the two states examined. In additional analysis, we have explored alternative thresholds of $a\times 1$ billion,  where $a=5,6,\ldots, 10$. We found that \$10 billion was the first threshold at which the regression analysis based on our proposed model revealed a statistically significant relationship between natural hazard losses in the large-loss regime and mortgage default rates.  This finding suggests that, based on our proposed model, for Texas and Louisiana, \$$10$ billion represents a statistically meaningful threshold for defining natural hazard shocks when viewed through the lens of mortgage default rates.}

\subsection{MCMC mixing}
\begin{figure}
    \centering
    \includegraphics[width=0.9\linewidth, height = 6cm]{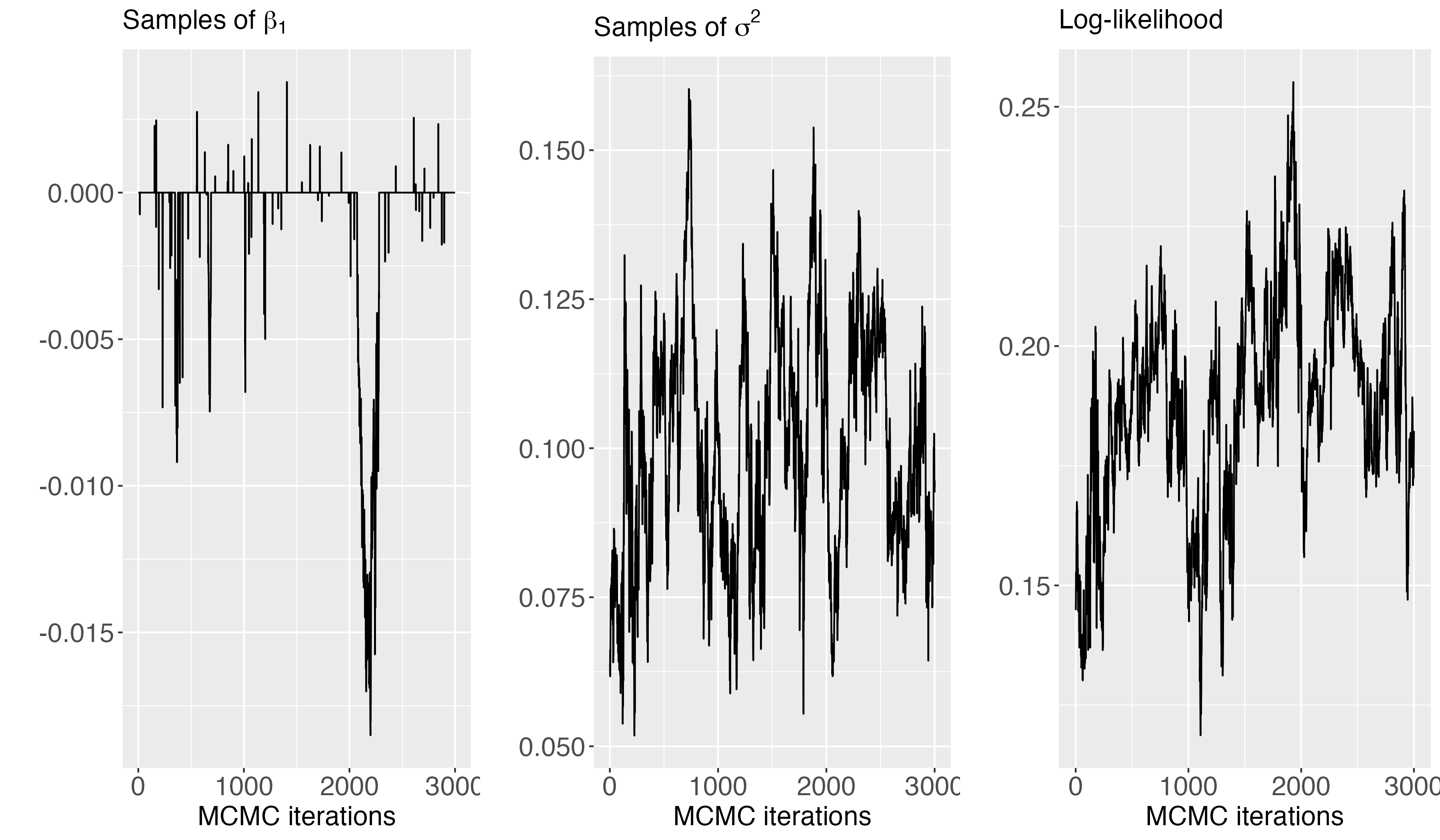}
    \caption{MCMC traceplots of $\beta_1$, $\sigma_y^2$, and $\ell$ representing the log-likelihood of the observed data for the {\bf SSM} model based on the Texas dataset. }
    \label{fig:texas_SSM_mixing}
\end{figure}

\begin{figure}
    \centering
    \includegraphics[width=0.9\linewidth, height = 6cm]{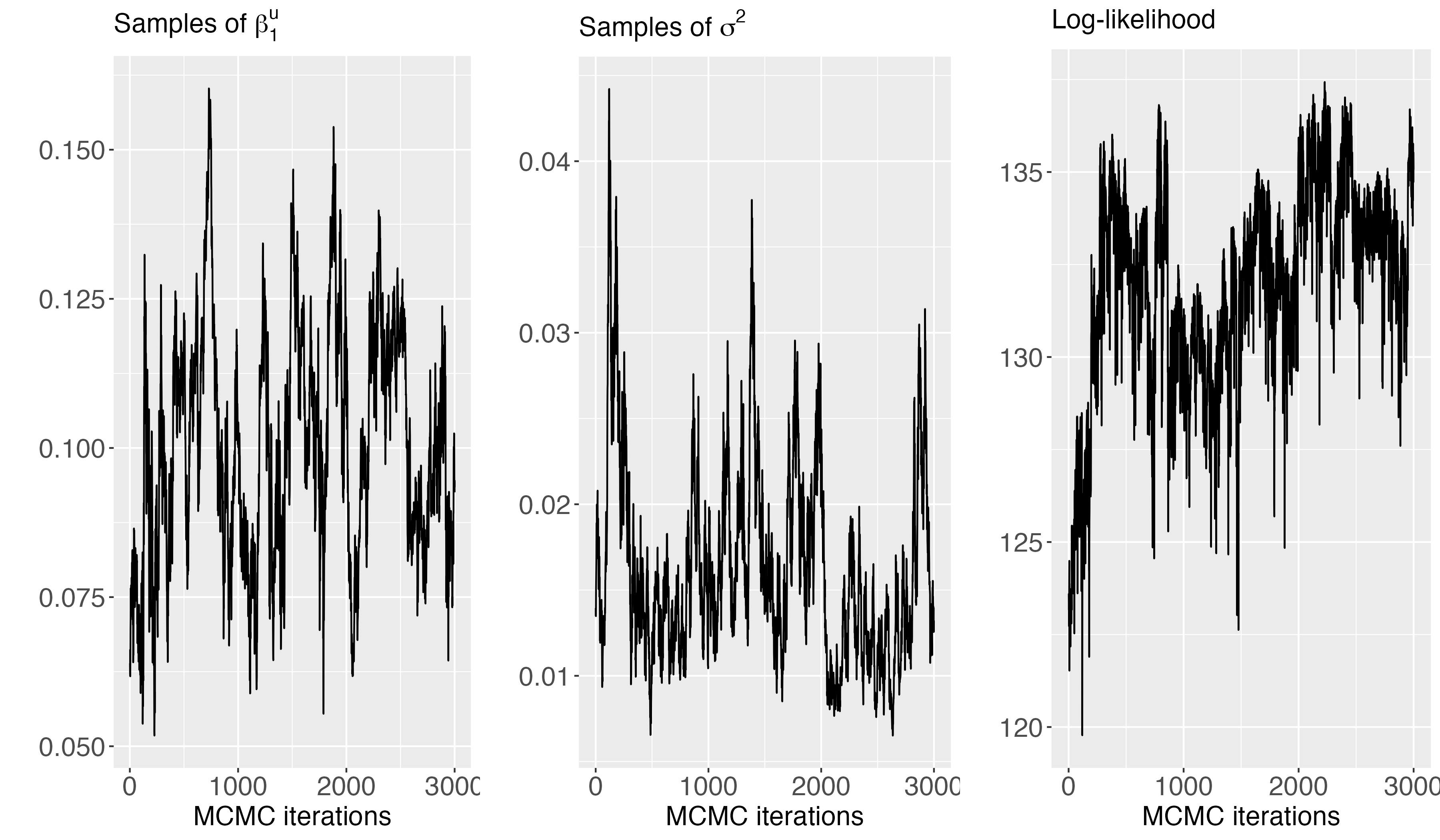}
    \caption{MCMC traceplots of $\beta_1$, $\sigma_y^2$, and $\ell$ representing the log-likelihood of the observed data for the {\bf mSSM} model based on the Texas dataset. }
    \label{fig:texas_mSSM_mixing}
\end{figure}

Latent variable models are prone to poor mixing of MCMC chains. Inferences drawn from poorly mixed chains can lead to erroneous conclusions. Hence, it is important to ensure that the MCMC chains have mixed reasonably before going into further analysis. Here, we first take a look at some descriptive quantities and visual evidences to ensure the chains have mixed. We first consider the {\bf SSM} and {\bf mSSM} model for the State of Texas. We ran the chain for $N= 5000$ iterations with 2000 burn-in iterations. In Figures \ref{fig:texas_SSM_mixing} and \ref{fig:texas_mSSM_mixing}, we show the traceplots of $(\beta_1, \sigma_y^2, \ell)$ and $(\beta_1^{u}, \sigma_y^2, \ell)$ for {\bf SSM} and {\bf mSSM}, respectively. Here, $\ell$ represents the observed log-likelihood. The chains seem to mix reasonably well; the log-likelihood of the observed data hovers around a central value. Another thing is apparent from these plots: for the {\bf SSM} model, where no distinction is made between very large and small natural hazard losses, the estimate of $\beta_1$ is very close to 0, in fact in almost half of the iterations it is set to 0. This matches with our intuition from Figure \ref{fig:outlier_loss_dlq} where there is no apparent (linear) relationship between natural hazard losses and default rates. For the {\bf mSSM} model however, we see that natural hazard events with catastrophic losses do impact the mortgage default rates. 

We also compute the effective sample size for each parameter. The effective sample size is roughly an estimate of how many ``independent" samples were generated out of the $N$ samples. For a generic parameter $\theta$, it is computed using the expression
$$N_{eff} = \dfrac{N}{1 + 2\sum_{k=1}^\infty r_k},$$
where $r_k$ is the lag-$k$ correlation of $\theta$. In practice, $k$ is set to some large number (30, in our case). For the {\bf SSM} model, the effective sample sizes for $N = 5000$ MCMC iterations for parameters $(\beta_1, \beta_2, \beta_3, \sigma_y^2, \sigma^2_\mu, \sigma^2_\nu)$ are (131, 470, 173, 89, 80, 81), respectively. The relatively low effective sample sizes might be due to the latent variable nature of the model. For the {\bf mSSM} model, the parameters are $(\beta^l, \beta^u, \sigma_y^2, \sigma^2_\mu, \sigma^2_\nu)$. The corresponding numbers are of the same order. We also computed results using different initializations, but generally results (mixing and estimates) were not sensitive to the choice of the initial value. 

\subsection{Inference on latent state $\mu_t$}
The posterior distribution of the latent variables indicate key differences between {\bf SSM} and {\bf mSSM}, especially for states that have large natural hazard losses exceeding the chosen threshold. As an example of such a state, we would again consider the state of Texas. Three major hurricanes, namely Hurricane Rita in 2005, Hurricane Ike in 2008, and Hurricane Harvey in 2017, contributed to large natural hazard losses. These events were followed by sharp spikes in delinquency rates; see left panel of Figure \ref{fig:texas-loss-dlq} for reference. Latent variables capturing the general macroeconomic features can not fully explain these spikes during these times. We will now see how this intuition is reflected in the actual results.
\begin{figure}
    \centering
    \includegraphics[width=0.8\linewidth, height = 5cm]{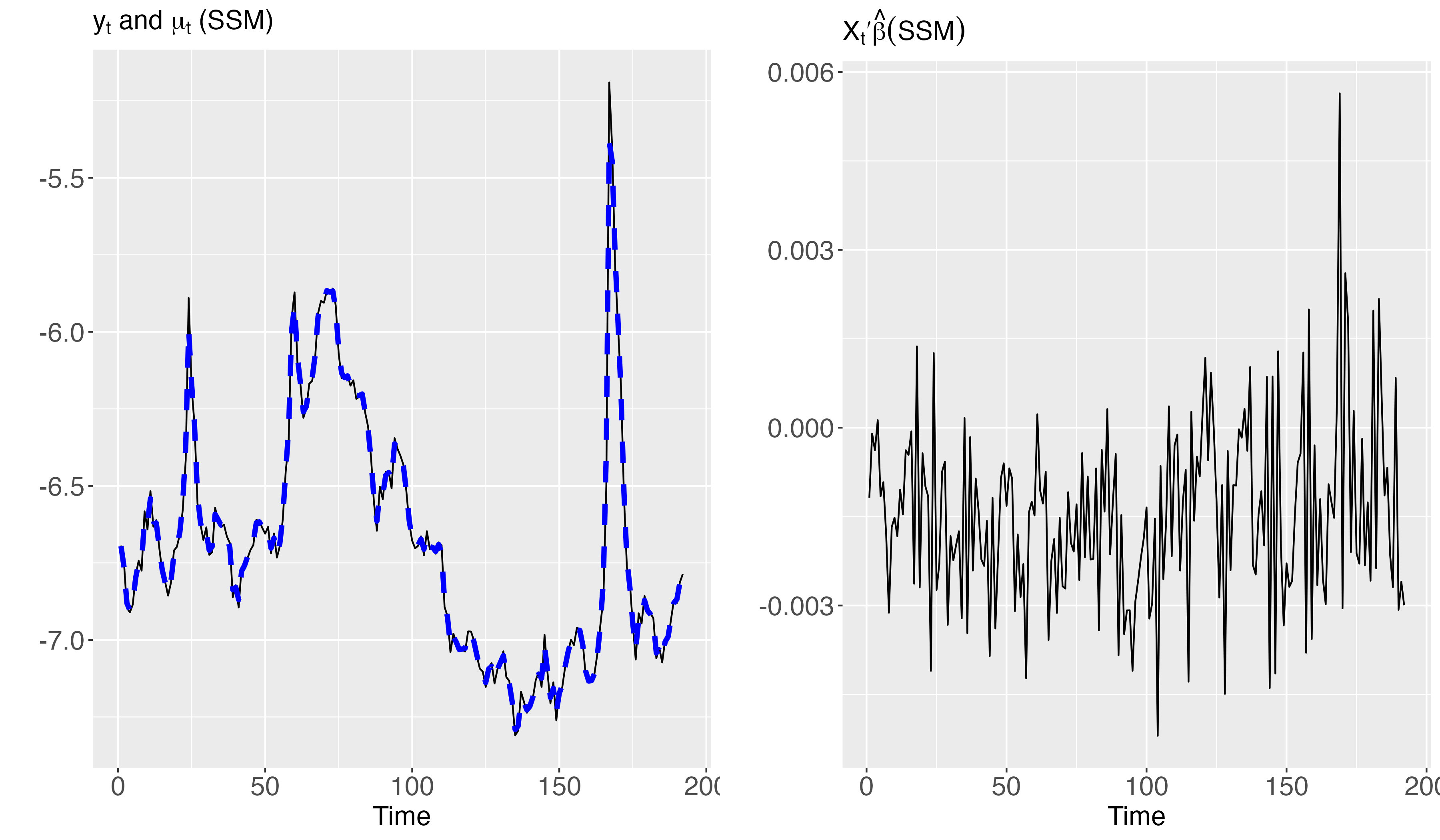}
    \caption{Posterior mean paths of the latent states $(\mu_t, \nu_t)$ for Texas obtained for the {\bf SSM} model. Black line is the observed data and blue dashed line is the posterior mean path of $\mu_t$.}
    \label{fig:texas_SSM_latent_states}
\end{figure}

\begin{figure}
    \centering
    \includegraphics[width=0.8\linewidth, height = 6cm]{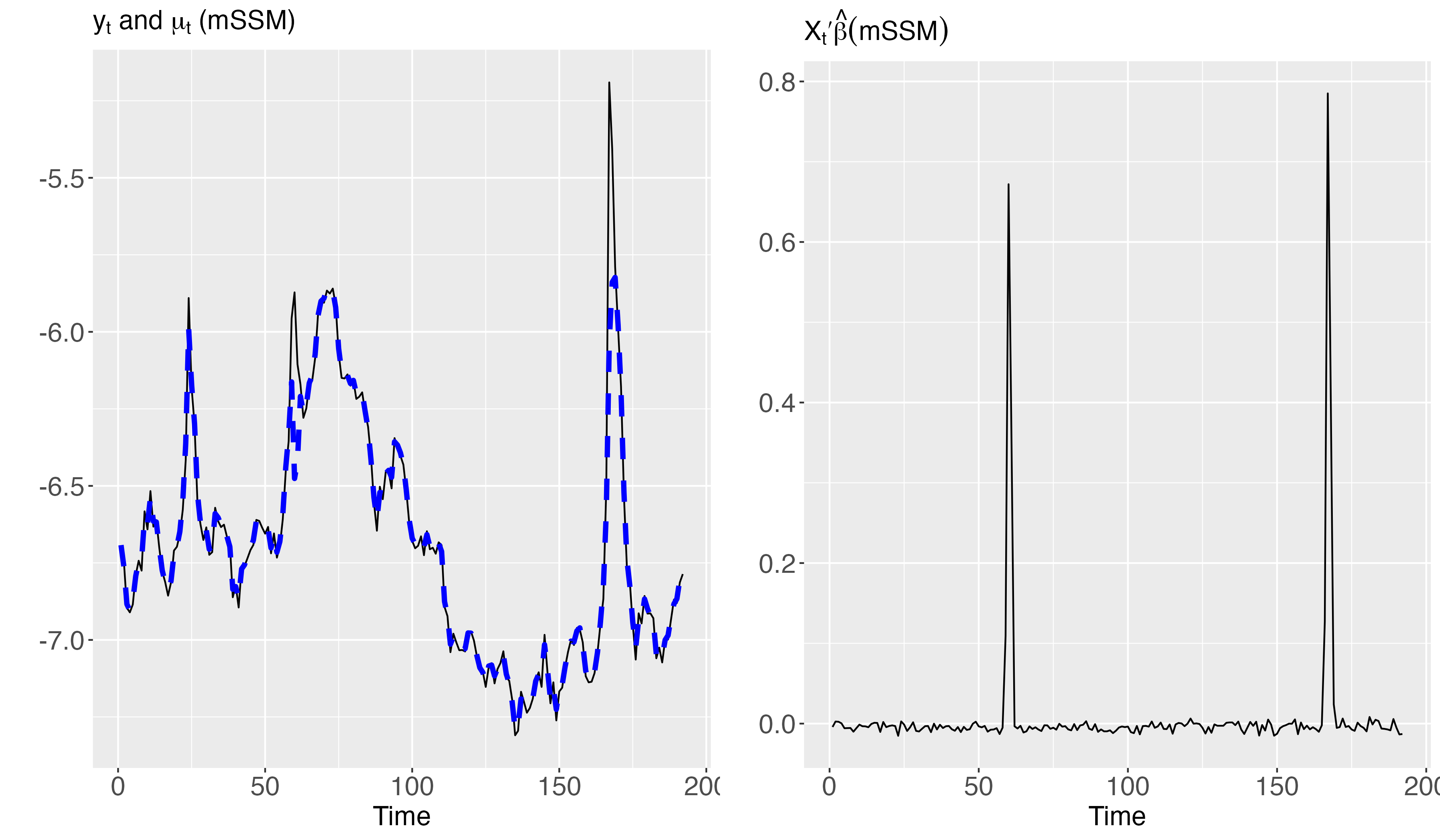}
    \caption{Posterior mean paths of the latent states $(\mu_t, \nu_t)$ for Texas obtained from the {\bf mSSM} model. Black line is the observed data and blue dashed line is the posterior mean path of $\mu_t$.}
    \label{fig:texas_mSSM_latent_states}
\end{figure}

The MCMC sampler with $N$ iterations provides us with $N$ samples of the path $(\alpha_t)_{t=1}^T$ given the observed data. From this we create the posterior mean path, which is simply the average over MCMC samples at each time point. Recall $\mu_t$ captures the state of the general economic system at time $t$, and $\nu_t$ captures the local trend, if any, at time $t$. In the left panels of Figure \ref{fig:texas_SSM_latent_states} and \ref{fig:texas_mSSM_latent_states}, we show the posterior mean path of $\mu_t$. The right panel of these figures show the regression component of these models, i.e. $X_t^\top \hat{\beta}$. It can be readily seen from these figures that for {\bf SSM}, $\mu_t$ almost identically follows the observed default rates. This is due to the lack of explanatory power of natural hazard losses as a whole on default rates - the regression component $X_t^\top \beta$ hovers around zero. On the other hand, for the {\bf mSSM} model, the situation is very different, in that $\mu_t$ falls short of capturing the amplitude of the spike in default rates. The remaining effect is then attributed to the catastrophic losses that happened just preceding the spikes. Additionally, $X_t^\top \beta$, increases sharply around the spikes in default rates. In summary, one can come to the following conclusions: 1) the general state-space model {\bf SSM} is a powerful model to understand the dynamics of default rates as governed by latent macroeconomic trends, and 2) if the objective is to study the effect of catastrophic losses on default rates, adjustments in the model specification such as \eqref{eq:modified_model} considered in this paper, are crucial to capture these effects. %In the Appendix, we show similar plots of other states that have experienced large catastrophic losses, such as ``Louisiana", and also compare them with states who have not experienced large losses.  

\subsection{Inference on $\beta^u$ and $\beta^l$}
\label{sec:inf-beta}
\begin{figure}
    \centering
    \includegraphics[width=0.95\linewidth, height = 6cm]{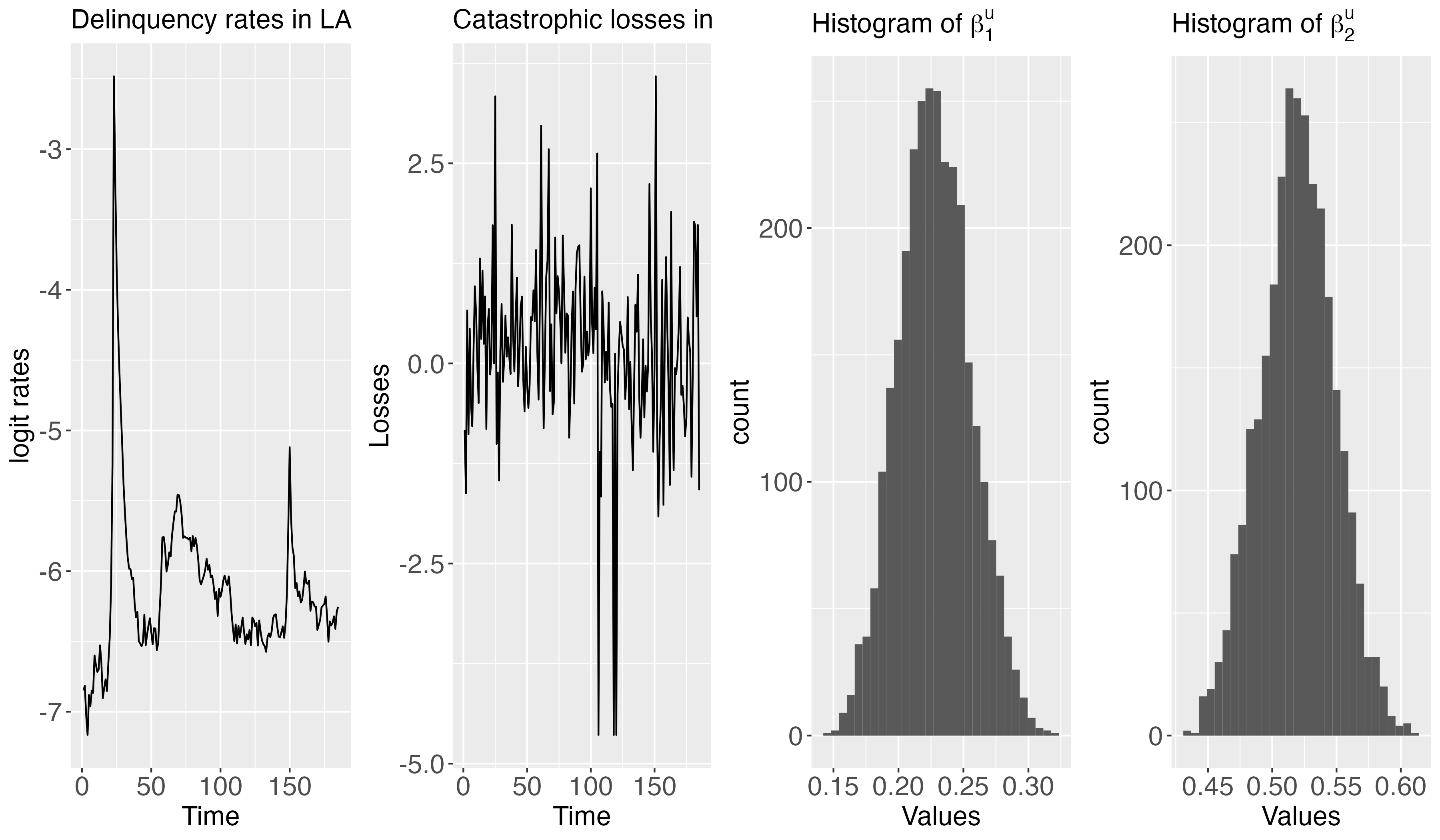}
    \caption{Analysis of Louisiana mortgage default rates. From left to right: default rates in logit scale, transformed natural hazard losses, histogram of posterior samples of $\beta_1^u$, histogram of posterior samples of $\beta_2^u$.}
    \label{fig:louisian_results}
\end{figure}

We discuss the estimates of $\beta^{u}$ and $\beta^l$ in this subsection for states with history of catastrophic losses. Specifically, we consider Louisiana along with Texas. Like Texas, we have default rate data for Louisiana from around 2005 until 2019. Hurricane Katrina in 2005 and Hurricane Harvey in 2017 made landfall in Louisiana inflicting devastating losses. We consider the {\bf mSSM} model. In Figure \ref{fig:louisian_results}, we show the default rates and the natural hazard losses for the state of Louisiana. As expected, there is a spike in default rates after Hurricane Katrina. This is also reflected in the posterior histograms of $\beta^u_1$ and $\beta^u_2$ shown in Figure \ref{fig:louisian_results}, where we can see that large losses \emph{increase} the default rates after removing the effect of latent macroeconomic factors. Indeed, the distribution of both $\beta^u_1$ and $\beta^u_2$ are almost entirely supported on the positive part of the real line. This is not the case with $\beta^l$. It is also important to note here that due to the prior structure on $\beta^u$ (spike and slab), if a predictor is deemed to have no effect on the default rates, then it will be set to zero in the posterior. We summarize these results for Texas and Louisiana in Tables \ref{tab:beta_posterior_summaries} and \ref{tab:low_beta_posterior_summaries} for Texas and Louisiana, respectively. Here we define
%We see that this is not the case for the first two components of $\beta^u$ although the third component is set to zero with positive probability. We give these details in Table \ref{tab:beta_posterior_summaries}, where we define 
\begin{align*}
\hat{\rho}^u_j = P[\beta^u_j \neq0 \mid y, X] = \frac{1}{N}\sum_{l=1}^N \mathbbm{1}(\beta^{u}_{j,l} \neq 0 ),  \quad \bar{\beta}^u_j = \frac{1}{N}\sum_{l=1}^N \beta^{u}_{j,l}, \quad 
 \text{sd}(\beta^u_j) = \frac{1}{N}\sum_{l=1}^N (\beta^{u}_{j,l} - \bar{\beta}^u_j)^2. 
\end{align*}
In the above display, $\beta_{j,l}^u$ is the $l$-th MCMC sample of $\beta^u_j$. We use the same definition and notation for posterior summaries of $\beta^l$. These quantities represent the posterior probability that $\beta^u_j$ is set to 0 (often called the posterior inclusion probability) and the posterior mean, standard deviation, respectively. Additionally, we define $\text{CI}(\beta^u_j)$ to be the 95\% credible interval obtained from the MCMC samples. 

As can be seen from the posterior inclusion probabilities of $\beta^u$, the first and second components, i.e. losses in months $t-3$ and $t-4$ are always included in the fitted model. The posterior distribution of $\beta^u_3$, which captures the effect of losses in month $t-5$, is a mixture --- it is non-zero with probability 0.58 for Texas and with probability 0.12 for Louisiana. The posterior means and quantiles also support this conclusion in that the 95\% credible interval for $\beta^u_1$ and $\beta^u_2$ are clearly shifted away from zero, which is not the case for $\beta^{u}_3$. Overall, these results seem to support our hypothesis that very large natural hazard losses induce a positive effect on mortgage default rates. Moreover, these same posterior summaries for $\beta^l$, reported in Table \ref{tab:low_beta_posterior_summaries}, show that the effects of natural hazard losses have a pronounced effect on delinquency rates only when these exceed a certain threshold. For example, none of the elements of $\beta^l$ are included in the model with probability more than 50\%. Their estimates are also very close to zero. This is true for both Texas and Louisiana.
\begin{table}[ht]
    \centering
    \def\arraystretch{1.3}
    \scalebox{0.9}{
    \begin{NiceTabular}{c|ccc|ccc}[hvlines]
    \toprule
     & \multicolumn{3}{c}{Texas} & \multicolumn{3}{c}{Louisiana} \\
    \midrule
     & $\beta^u_1$ & $\beta^u_2$ & $\beta^u_3$ & $\beta^u_1$ & $\beta^u_2$ & $\beta^u_3$ \\
    \midrule
    $\hat{\rho}$      & 1    & 1    & 0.58  & 1    & 1    & 0.12 \\
    $\bar{\beta}$     & 0.09 & 0.17 & 0.02  & 0.23 & 0.52 & 0.006 \\
    $\text{sd}(\beta)$ & 0.03 & 0.02 & 0.02  & 0.03 & 0.03 & 0.01 \\
    $\text{CI}(\beta)$ & (0.05, 0.14) & (0.13, 0.22) & (0.00, 0.08) & (0.17, 0.28) & (0.46, 0.58) & (0.00, 0.04) \\
    \bottomrule
    \end{NiceTabular}
    }
    \caption{Posterior summaries of $\beta^u$ for Texas and Louisiana obtained from the {\bf mSSM} model.}
    \label{tab:beta_posterior_summaries}
\end{table}

\begin{table}[ht]
    \centering
    \def\arraystretch{1.3}
    \scalebox{0.9}{
    \begin{NiceTabular}{c|ccc|ccc}[hvlines]
    \toprule
     & \multicolumn{3}{c}{Texas} & \multicolumn{3}{c}{Louisiana} \\
    \midrule
     & $\beta^l_1$ & $\beta^l_2$ & $\beta^l_3$ & $\beta^l_1$ & $\beta^l_2$ & $\beta^l_3$ \\
    \midrule
    $\hat{\rho}$      & 0.07    & 0.11    & 0.32  & 0.03   & 0.22    & 0.05 \\
    $\bar{\beta}$     & $6\times 10^{-5}$  & $9 \times 10^{-4}$ & $-5\times 10^{-3}$ & $1\times 10^{-4}$ & $3 \times 10^{-3}$ & $-3\times 10^{-4}$ \\
    $\text{sd}(\beta)$ & 0.002 & 0.003 & 0.009  & 0.001 & 0.006 & 0.002 \\
    $\text{CI}(\beta)$ & (-0.003, 0.003) & (0.00, 0.01) & (-0.03, 0.00) & (0.00, 0.00) & (0.00, 0.02) & (-0.007, 0.00) \\
    \bottomrule
    \end{NiceTabular}
    }
    \caption{Posterior summaries of $\beta^l$ for Texas and Louisiana obtained from the {\bf mSSM} model.}
    \label{tab:low_beta_posterior_summaries}
\end{table}
\subsection{Prediction}
We next focus on the task of prediction. For this, we shall consider a rolling-window prediction in the following sense. Suppose $(y_t, X_t)_{t=1}^{t_0}$ represent the $t_0$ points of observations (for any one state). We use this data to predict $y_{t_0+1}$ using the models {\bf SSM} and {\bf mSSM}. To predict $y$ at $t_0+2$, we use $(y_t, X_t)_{t=1}^{t_0+1}$ to train the model, so on and so forth. Both of these models allow us to compute the posterior predictive distributions defined as
\begin{align*}
p(y_{t_0+1} \mid (y_t, X_t)_{t=1}^{t_0} ) &= \int p(y_{t_0+1} \mid \alpha_{t_0+1}, \theta) p(\alpha_{t_0+1}, \theta\mid (y_t, X_t)_{t=1}^{t_0} ) d\alpha_{t_0+1} d\theta\\
 = \int & p(y_{t_0+1} \mid \alpha_{t_0+1}, \theta) p(\alpha_{t_0+1} \mid \alpha_{t_0}, \theta) p(\alpha_{t_0}, \theta \mid (y_t, X_t)_{t=1}^{t_0})d\alpha_{t_0+1} d\alpha_{t_0} d\theta.
\end{align*}
In other words, a sample from $p(y_{t_0+1} \mid (y_t, X_t)_{t=1}^{t_0} )$ can be drawn by first sampling $\alpha_{t_0}, \theta \mid (y_t, X_t)_{t=1}^{t_0}$, then $\alpha_{t_0+1} \mid \alpha_{t_0}, \theta$ and finally drawing $y_{t_0+1} \mid \alpha_{t+1}, \theta$. Samples of $\alpha_{t_0}, \theta \mid (y_t, X_t)_{t=1}^{t_0}$ are already available from the MCMC implementation, and the distribution of $\alpha_{t_0+1}$ and $y_{t_0+1}$ are available from the model \eqref{eq:model}. 

A key distinction in this case from classical frequentist inference is that we obtain a predictive distribution instead of a point forecast. In order to evaluate this predictive distribution against point realizations, we will use scoring rules \citep{gneiting2007strictly}. Specifically, we will use the Continuous Ranked Probability Score (CRPS). The advantage of using this scoring rule is that it takes into account the calibration of the entire predictive distribution against the observed value, instead of simply focusing on some location of the predictive distribution, such as its mean or median. Suppose $F(x)$ is the cumulative predictive distribution obtained by some method, and let $y$ be the observed value. Then the CRPS is defined as
$$\text{CRPS}(F, y) = \int \{F(x) - \mathbbm{1}_{(x\geq y)}\}^2 dx,$$
where $\mathbbm{1}_{(x\geq y)} = 1$ if $x \geq y$ and $0$ otherwise. Intuitively, $\mathbbm{1}_{(x\geq y)}$ is the cumulative distribution function of a degenerate random variable taking the value $y$. This is compared with the forecast distribution $F(x)$. When a point forecast $\hat{y}$ is available instead of a distribution, the CRPS reduces to the absolute error. In our computation, since the mortgage rates in the original scale are numbers between 0 and 1, we approximate the CRPS score by numerical integration using a quadrature method:

$$\text{CRPS}(F, y) \approx \sum_{k=1}^m w_k \{\hat{F}(v_k) - \mathbbm{1}_{(v_k\geq y)}\}^2,$$
where $(v_k, w_k)_{k=1}^m$ represent the $m$ quadrature points within the interval $[0,1]$, and $\hat{F}$ is the empirical predictive distribution function constructed from the posterior predictive samples.

We focus on the state of Texas for which we have $T = 192$ observations and compare the CRPS scores for {\bf SSM} and {\bf mSSM}. For the rolling-window, we consider $t_0 = 150,\ldots,  191$, and computed the average CRPS score
We obtain that the CRPS scores for {\bf SSM} and {\bf mSSM} are $3.9\times 10^{-7}$ and $1.7 \times 10^{-8}$, respectively. Thus, {\bf mSSM} provides an improvement of an order of magnitude over {\bf SSM} in terms of CRPS. Naturally, for states which do not encounter large natural hazard losses, CRPS scores of {\bf SSM} are {\bf mSSM} are almost exactly equal. 
\begin{figure}
    \begin{subfigure}{0.5\textwidth}
        \centering
        \includegraphics[width = 7cm, height = 7cm]{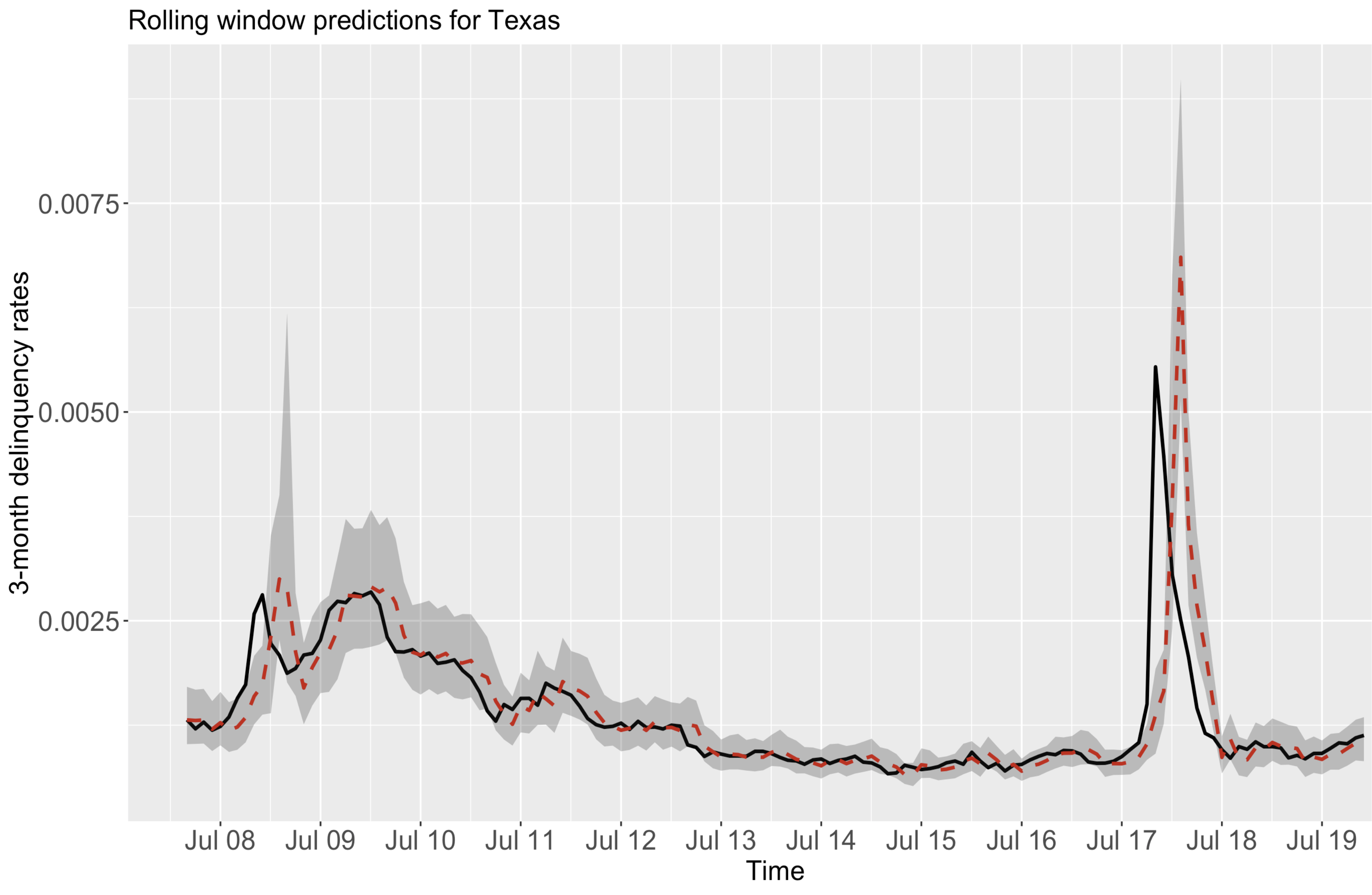}
    \end{subfigure}
    \begin{subfigure}{0.5\textwidth}
        \centering
        \includegraphics[width = 7cm, height = 7cm]{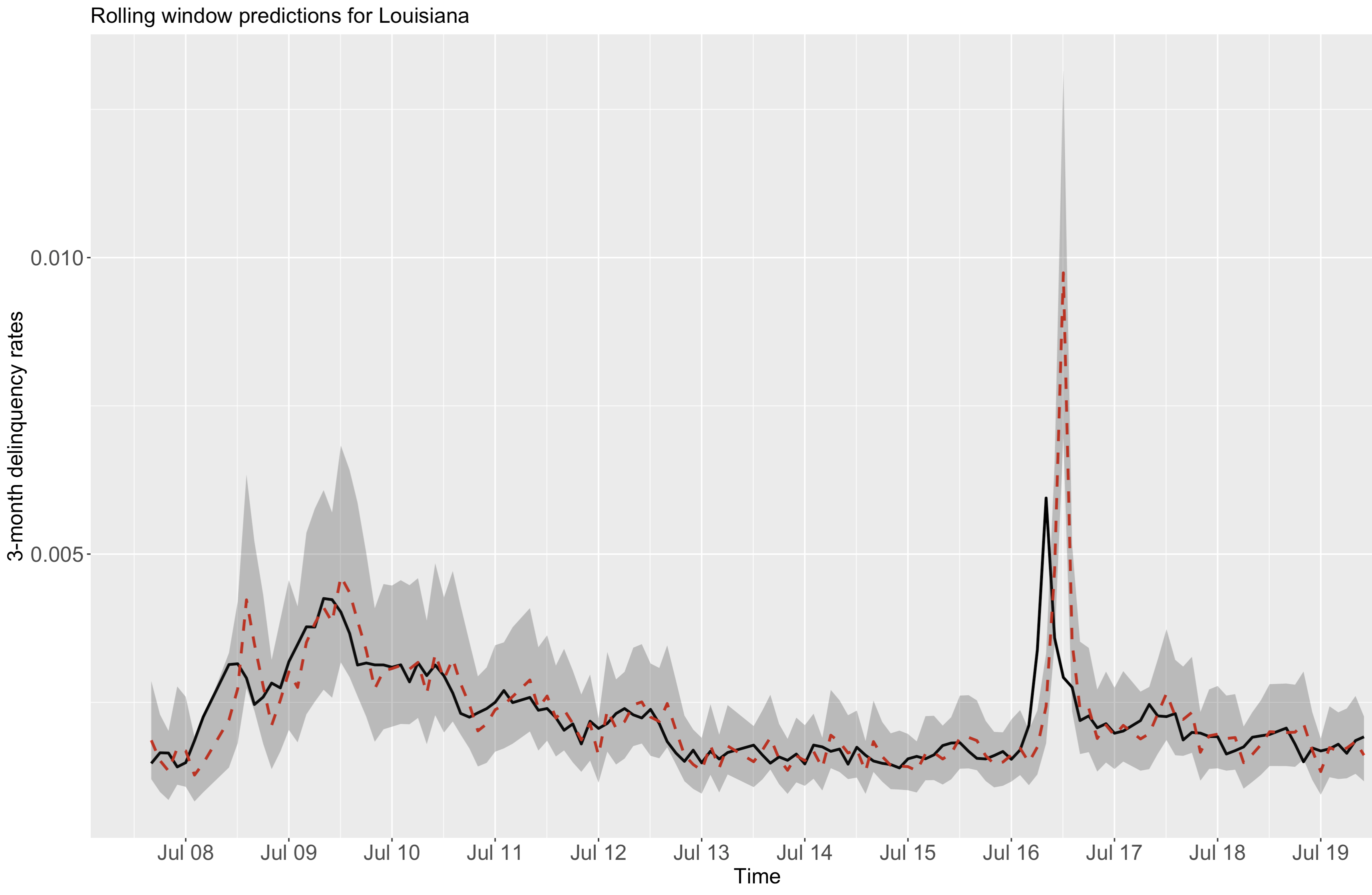}
    \end{subfigure}
    \caption{Rolling window predictions for Texas and Louisiana obtained from {\bf mSSM}. Here, large spikes around the years 2017 in Texas and  Louisiana correspond to Hurricane Harvey. Black solid lines depict the observed delinquency rate, red dashed lines trace the posterior mean of the predictive distribution, and gray shaded region shows the 95\% empirical credible interval of the predictive distribution. }
    \label{fig:TX_LA_predictions}
\end{figure}

\begin{figure}
    \begin{subfigure}{0.42\textwidth}
        \centering
        \includegraphics[width=7cm, height=6cm]{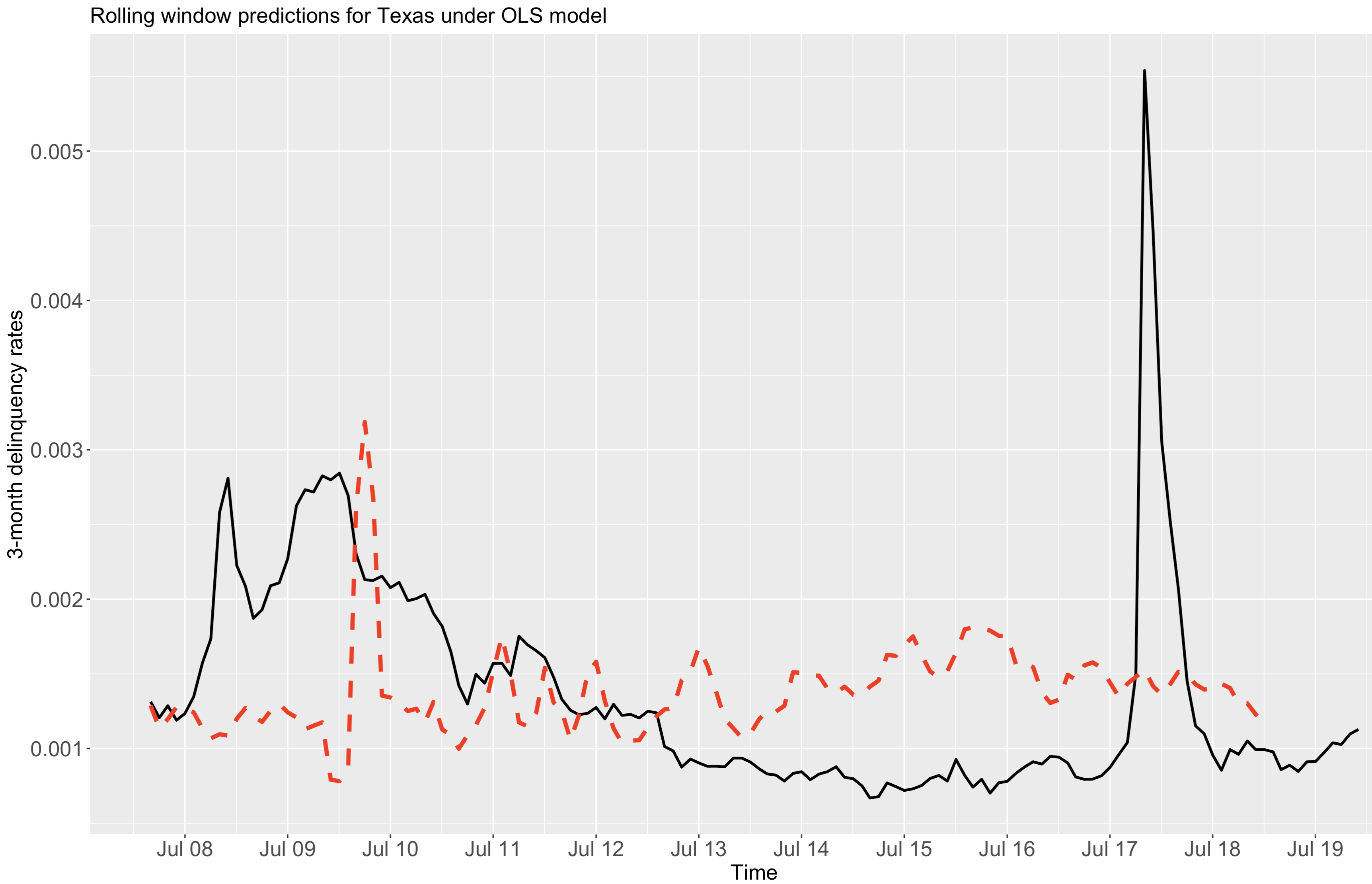}
    \end{subfigure}
    \qquad
    \qquad
    \begin{subfigure}{0.42\textwidth}
        \centering
        \includegraphics[width=7cm, height=6cm]{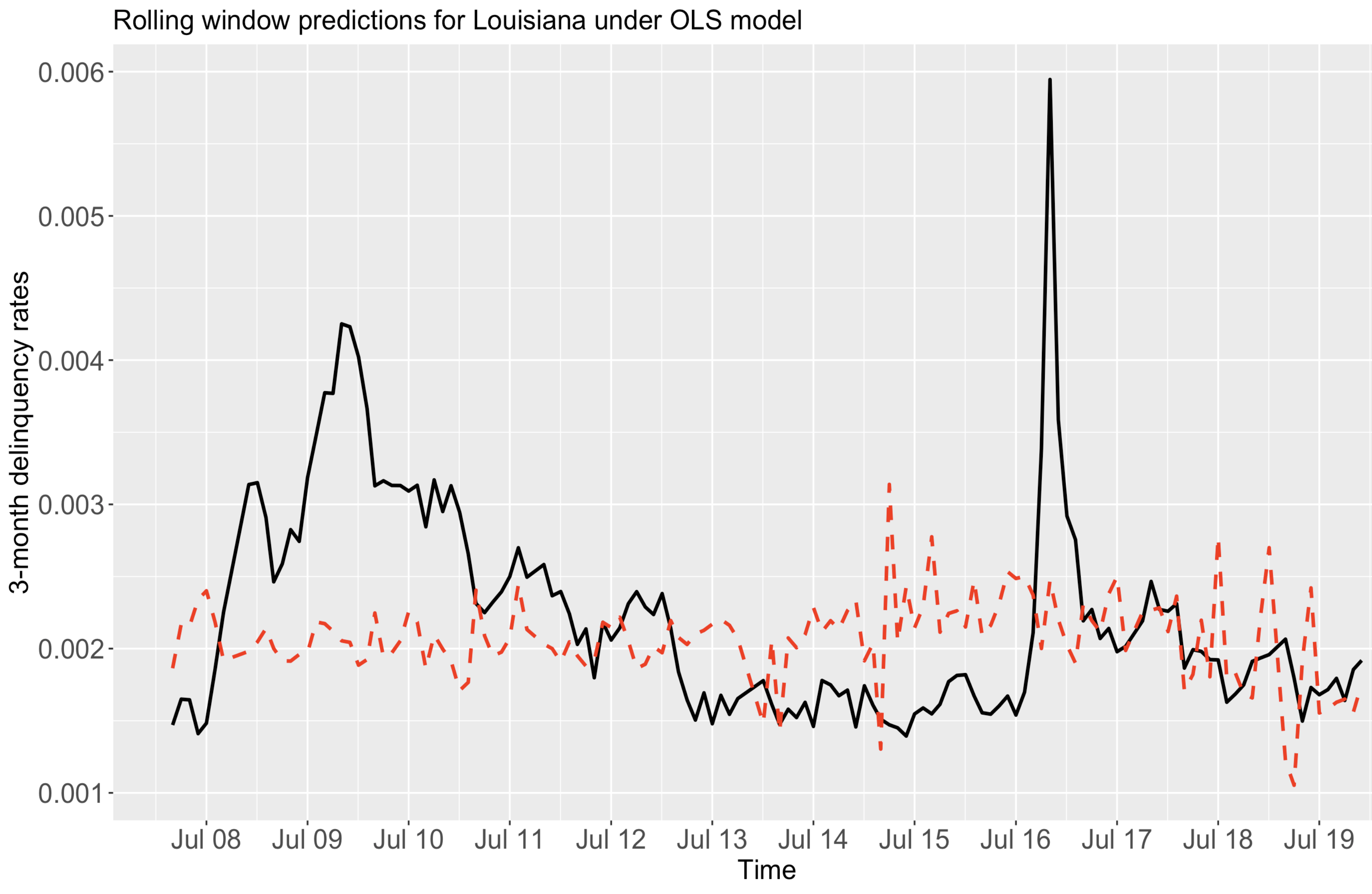}
    \end{subfigure}
    \caption{One-step-ahead predictions of mortgage default rates for Texas and Louisiana based on a simple OLS model without latent states. Black line corresponds to observed mortgage default rates whereas the red dashed line shows one-step predictions.}
    \label{fig:texas_ols}
\end{figure}

In Figure \ref{fig:TX_LA_predictions}, the rolling window predictions in the original scale are shown for Texas and Louisiana for {\bf mSSM}. Here, we set $t_0 = 50$. We can see that the proposed model is able to predict large spikes anticipated from huge loss inflicted by natural hazards.
{\dr We are not aware of any earlier work that explicitly examines the impact of natural hazard losses on increases in mortgage default rates; therefore, we do not have a directly ``existing'' model for comparison. In practice, ordinary least squares (OLS) regression is a simple yet widely used approach for studying relationships between explanatory and response variables in panel or longitudinal data.
Given the absence of a prior model, we use OLS linear regression as a naive benchmark to illustrate the benefits of adopting a SSM for capturing unobserved dynamics beyond the direct effects of natural hazard shocks on mortgage default rates. This benchmark model is identical to equation \eqref{eq:model} in our paper, except that it excludes the random effect term. Figure \ref{fig:texas_ols} displays the corresponding prediction results based on the simple OLS regression approach. When compared to the prediction performance shown in Figure \ref{fig:TX_LA_predictions}, the advantages of incorporating the SSM framework become readily apparent.}

\section{Conclusion}
\label{sec:conclusion} 
In this article, we study the relationship between natural hazard losses and mortgage default risks. Our exploratory analysis on the Fannie Mae default data and SHELDUS natural hazard loss data show empirical evidence of the impact of natural hazard losses on default risks. We formulate a sliced SSM to formally model default risks with natural hazard losses. The proposed model allows very large losses to impact differently on default risks through a careful choice of coefficients. Results of our analysis on data from the states of Texas and Louisiana show default rates can be positively effected by large natural hazard losses from the 3 and 4 months prior. 

As for future work, we envision natural extension of this model to consider non-linear effects of natural hazard loss data through nonparametric Bayesian priors for functions. A separate line of research may potentially consider a hierarchical model for default data from all the states, ensuring borrowing of information between the states. This modeling approach has the benefit of informing inference on states who have not experienced large losses through similar effects on states who have experienced these losses. 

\section*{Acknowledgment}
{\dr We sincerely thank the Editor and two anonymous Reviewers for their thoughtful and constructive feedback, which have significantly improved the clarity, rigor, and overall quality of our paper.}
We gratefully acknowledge the generous financial support provided by the Society of Actuaries for this project through the Individual Grant.

%\newpage
\bibliography{refs}
\bibliographystyle{chicago}
%\newpage
\appendix

\end{document}